\documentclass[10pt,a4paper]{article}

\usepackage[latin1]{inputenc}
\usepackage{amsthm,amsmath,amsfonts} 
\usepackage{amssymb}
\usepackage{mathrsfs}

\usepackage{eurosym}
\usepackage{verbatim} 
\usepackage[english]{babel}
\usepackage[T1]{fontenc}
\usepackage{bbm,bm}

\usepackage{lmodern}
\usepackage{blindtext}

\usepackage{tabularx}
\usepackage[table,xcdraw]{xcolor}
\usepackage{graphicx, floatflt, subfig}
\usepackage{color}

\usepackage{multirow}
\usepackage{booktabs}
\usepackage{adjustbox}
\usepackage{blindtext}
\usepackage{esint}
\usepackage[toc]{appendix}

\theoremstyle{definition}
\newtheorem{theorem}{Theorem}[section]
\newtheorem{lemma}{Lemma}[section]

\newtheorem{definition}{Definition}[section]

\begin{document}
\title{\bf Global and Tail Dependence: \\A Differential Geometry Approach   }
\author{
	Davide Lauria\thanks{Texas Tech University, Department of Mathematics
		\& Statistics, Lubbock TX 79409-1042, U.S.A., Davide.Lauria@ttu.edu}
	\and 
	Svetlozar T. Rachev\thanks{Texas Tech University, Department of Mathematics
		\& Statistics, Lubbock TX 79409-1042, U.S.A., Zari.Rachev@ttu.edu }	
	\and 
	A. Alexandre Trindade\thanks{Texas Tech University, Department of Mathematics
		\& Statistics, Lubbock TX 79409-1042, U.S.A., Alex.Trindade@ttu.edu.}	
}
\maketitle

\begin{abstract}
Measures of tail dependence  between random variables aim to  numerically quantify the degree of association between their extreme realizations. Existing tail dependence coefficients (TDCs) are based on an asymptotic analysis of relevant conditional probabilities, and do not provide a complete framework in which to compare extreme dependence between two random variables. In fact, for many important classes of bivariate distributions, these coefficients take on non-informative boundary values. We propose a new approach by first considering global measures based on the surface area of the conditional cumulative probability in copula space, normalized with respect to departures from independence and scaled by the difference between the two boundary copulas of co-monotonicity and counter-monotonicity. The measures could be approached by cumulating probability on either the lower
left or upper right domain of the copula space, and offer the novel perspective of being able to differentiate asymmetric dependence with respect to direction of conditioning. The resulting TDCs
produce a smoother and more refined taxonomy of tail dependence. The empirical performance of the measures is examined in a simulated data context, and illustrated through a case study examining tail
dependence between stock indices.
\end{abstract}

\providecommand{\keywords}[1]
{
  \small	
  \textbf{\textit{Tail dependence}} #1
}


\section*{Introduction}\label{sec:intro}

The analysis of the extremal dependence between two random variables takes on great importance in a wide range of scientific disciplines and applications like computer science, environmental engineering, meteorology, risk management, and finance.  
\textit{Tail dependence coefficients} (TDCs) between two random
variables $X$ and $Y$, with joint distribution function $H$ and
marginal distribution functions $F$ and $G$ respectively, were introduced by \cite{Sibuya1960}  in order to measure the dependence between their extreme realizations.
The lower TDC is a measure of the dependence in the lower-left quadrant and is defined as:
\begin{align}
   \label{Sibuya_l}
   & \lambda_{l}(X|Y) = \lim_{\alpha \to 0^{+} } \mathbb{P}\left[ X \le F^{-1} \left( \alpha \right) | Y \le G^{-1} \left( \alpha \right) \right ] = \lim_{\alpha \to 0^{+} } \frac{ C(u,v) }{ v }.
\end{align}
Analogously, the upper TDC measures dependence in the upper-right
quadrant as:
\begin{align}
   \label{Sibuya_u}
   & \lambda_{u}(X|Y) = \lim_{\alpha \to 1^{-} } \mathbb{P}\left[ X > F^{-1} \left( \alpha \right)  | Y> G^{-1} \left( \alpha \right) \right]= \lim_{\alpha \to 1^{-} } \frac{ 1 - u - v + C(u,v) }{ 1-v } .
\end{align}
Note that Sklar's theorem (Theorem~\ref{th:Sklar} in
Appendix~\ref{copulatheory})  immediately implies that  $\lambda_{l}$
and $\lambda_{u}$ are completely determined by the implied copula $C$
that connects $H$, $F$, and $G$. Both measures are confined to the interval $[0,1]$, where a zero value denotes tail independence\footnote{They
have been generalized to random vectors of dimensions greater than
two; see \cite{joe1997multivariate} for a review.}.

TDCs, also known as \emph{strong} TDCs, are intuitively simple to
understand, but suffer from important theoretical and practical
drawbacks. In particular, two main limitations have been pointed out:
(i) the coefficients do not convey any information on the rate of
convergence of the limit in (\ref{Sibuya_l})--(\ref{Sibuya_u}), and
(ii) the dependence is measured only along the diagonal $x=y$. The consequence, from a practical perspective, is that TDCs assume their boundary values, zero or one, for many popular  distribution classes, leading to a weak taxonomy of the behavior of the conditional structure for joint extreme events. 

The case of elliptically contoured distributions provides an illustrative example.  
An $n$-dimensional random vector $Z$ is called elliptically distributed
with parameters  $\mu \in \mathbb{R}^{n}$  and $\Sigma \in
\mathbb{R}^{n \times n}$ if $Z\overset{d}{=} \mu + R_{n}A'U$, where $A
\in \mathbb{R}^{n \times n}$,  with $A'A=\Sigma$ and rank$\left(
  \Sigma \right) = n$. $R_{n}$ is a random variable with distribution
$F_{n}$, called the generating distribution, and $U$ is an
$n$-dimensional random vector uniformly  distributed on the unit
sphere, and independent of $R_{n}$. The random variable $E \overset{d}{=} R_{n}U$ is then called spherically distributed, and each margin of $Z$ has the same distribution $G$.  
Schmidt \cite{Schmidt_2002} proved that if $F_{n}$ has a regularly varying tail, then all bivariate margins of $Z$ are tail dependent (the TDCs are strictly positive); furthermore, the TDCs  of any bivariate marginal extracted from an elliptically contoured distribution are greater than zero  if the survival function $\bar{G}$ of the random variable $Y$ is $O$-regularly varying; if $G$ has instead a regularly varying tail, then  all  bivariate margins have tail dependence.  
Multivariate normal, logistic, and  symmetric generalized hyperbolic
distributions, for instance, are elliptical distributions with
bivariate margins which are tail independent.  A multivariate
$t$-distribution with $\nu$ degree of freedom and linear correlation coefficient $\rho$, has instead positive TDCs given by 
\begin{align}
\lambda_{l} = \lambda_{u} = 2t_{\nu+1}\left(  -\sqrt{ \left( \frac{(\nu+1)(1-\rho)}{1 + \rho} \right) } \right). \label{TDC_tstudent}
\end{align}
where  $t_{\nu}(x)$ is the density function of the univariate $t$ with $\nu$ degrees of freedom.

Another important example is given by the class of multivariate
\textit{generalized hyperbolic} (GH) distributions introduced in
\cite{BN77}. GH distributions play an important role in financial
modeling, in particular for their flexibility in capturing particular
stylized facts like asymmetries and fatter tails with respect to the
Gaussian case \cite{Aas_2006, Barndorff_1997, Eberlein_98,
  Bibby_2003, Hitaj_2013, wang_2009}. Hammerstein, \cite[Theorem
4]{Hammerstein16}, showed that TDCs for bivariate margins extracted
from multivariate GH distributions can only assume the values of zero
or one; the only exception being the multivariate  scaled and shifted
$t$-distribution (which has equal lower and upper coefficients). A
brief summary of bivariate GH distributions can be found in  Appendix~\ref{AppendixC}.

Again, as for elliptical distributions,  this result is in contrast to
the evidence that dependence structures in the lower-left and
upper-right quadrants can be  substantially greater than the
independent case, and also for parameter settings that produce TDCs
equal to zero. Furthermore, these extreme dependence relationships can
assume a wide range of behaviours.
Consequently, classification of tail dependence for these kinds of distributions is poor, and with limited practical application. Indeed, even in the bivariate Gaussian case when  both marginals have light-tails, it is evident that the degree of tail dependence in the west-lower (east-upper) corner of the support changes considerably as the correlation coefficient increases toward one. 


The first attempt at improving TDCs measured the speed at which the
conditional probability varies in the joint tails. \cite{Ledford1996}
based their analysis on the idea that the joint distribution $H$ is in
the domain of attraction of some multivariate extreme value
distribution. Restricting attention to the two-dimensional
case of interest in our paper, they first considered transformations
\begin{align} \label{FrechetTransform}
&Z_{1} = \frac{-1}{\log F(X)} && \text{ and } & Z_{2} =\frac{-1}{\log G(Y)}, 
\end{align}
so that the joint distribution $H_{Z}$ of $\left[Z_{1}, Z_{2} \right]$
satisfies $H_{Z}(z_{1},z_{2}) = H(x,y)$, where $Z_{1}$ and $Z_{2}$ follow a Fr\'echet distribution, i.e., $P(Z_{i} \le z) =e^{-1/z}$, for $i=1,2$. The behaviour of the joint distribution is then modelled by assuming the following asymptotic result
\begin{equation} \label{SlowingVarying_Frechet}
P(Z_{1} >z, Z_{2}>z) \sim \ell(z) z^{-\frac{1}{\eta}}, \ \ \ z \to \infty,
\end{equation} 
where $\ell(z)$ is a slowly varying function and $1/2\le\eta\le 1$. It
can be shown that if $\eta = 1$ and $\lim_{z \to \infty
}\ell(z)=\lambda>0$, then $X$ and $Y$ are upper tail dependent with
$\lambda_{u}=\lambda$. If instead  $\lim_{z \to \infty
}\ell(z)=0$, then $ \lambda_{u} =0 $ and $\eta<1$. In other words, if
$\ell(z)$ converges to zero, we do not have asymptotic dependence since the upper TDC is equal to zero. (The value of $\eta$ can be interpreted as the rate at which the dependence decreases in the joint upper tail
\footnote{The idea was then further developed  in \cite{Hua_2011}, where concepts of tail order (the ratio $\frac{1}{\eta}$) and tail order function have been introduced, see also paragraph 2.16 in \cite{joe1997multivariate}.}.) 

A strictly connected measure  is the \emph{weak} coefficient of tail dependence $\chi$, defined by \cite{Coles1999} as 
\begin{equation}\label{LWeakTD} 
\chi_{l} = \lim_{\alpha \to 0^{+} }  \frac{ \log{\left[ P(U \le \alpha
      )P(V \le \alpha) \right]} }{  \log{\left[  P(U \le \alpha,V \le
      \alpha)\right]} } - 1=\lim_{\alpha \to 0^{+} }  \frac{ 2
  \log{\alpha} }{ \log{ \left[ C(\alpha,\alpha) \right] } } - 1,
\end{equation}
for the lower tail, and 
\begin{equation}\label{UWeakTD} 
\chi_{u}  = \lim_{ \alpha \to 1^{-}} \frac{ \log{\left[ P(U>\alpha)P(V>\alpha) \right]} }{  \log{\left[  P(U>\alpha,V>\alpha)\right]} } - 1 = \lim_{\alpha \to 1^{-} }  \frac{ 2 \log{ \left(  1-\alpha \right) } }{ \log{ \left[ 1 - 2\alpha +C(\alpha,\alpha) \right]} } -1, 
\end{equation}
for the upper tail, where $U = F^{-1}(X)$ and
$V=G^{-1}(Y)$. Curiously, the coefficient
$\eta$ in \eqref{SlowingVarying_Frechet} is related to $\chi$ through
the equation $\chi=2\eta-1$, meaning that computation of the two
coefficients  are one and the same. This property has been exploited
in order to compute $\chi$ for particular classes of multivariate
distributions. \cite{Schlueter2012} for instance, obtain that
$\chi_{l}=\chi_{u}=\sqrt{2(1+\rho)}-1$ for the class of elliptical GH
distributions with $\rho\ge 0$, where $\rho$ is the correlation
coefficient between $X$ and $Y$, whereas in the Gaussian special case
we have instead $\chi_{l}=\chi_{u}=\rho$.
The result is encouraging since it means $\chi$
contains information on the rate at which the conditional probability
changes as we move either southwest (lower) or  northeast (upper)  on
the support. It is therefore  able to detect positive tail
dependence also in the case of semi-heavy marginal tails, offering
the possibility of potential improvements with respect to
TDCs. (\cite{Heffernan2000} compute the values of $\chi$ for various well known copulas.)

At this point we have two measures of tail dependence: the weak
($\chi$) and the strong ($\lambda$). The former is applicable in
detecting ``weak'' tail dependence  structures (meaning that 
they disappear asymptotically); the latter is instead better
suited for assessing asymptotic tail dependence.
A potential drawback, as pointed out by  \cite{Furman2015}, among
others, is that all these measures compute tail
dependence by focusing only on the main diagonal $u=v$ of the copula support.
The implications of this are possible under-estimation of the joint
tail risk \cite{Furman2016}. In order to avoid this limitation,
\cite{Furman2015} proposed reformulating weak TDCs by moving along the
path of maximal tail dependence. That is, both weak and strong TDCs
should be computed along the path where the copula assumes its largest values on its way toward $(0,0)$ (for lower tail dependence), and toward $(1,1)$ (for upper tail dependence).

In empirical finance, both strong and weak TDCs are often computed via
nonparametric methods based on extreme value theory. An example is the
estimator proposed by \cite{Hill_2010}, which is directly applied to
(\ref{SlowingVarying_Frechet}) in order to estimate $\eta$. Other
nonparametric estimators of tail dependence can be found in
\cite{Schmidt_2006}, \cite{Salazar_2015}, and
\cite{Ferreira_2015}. These methods have the advantage of avoiding the
aforementioned limitation in the TDCs, since they do not rely on
direct computation of the limits with respect to a particular
copula. However, nonparametric approaches usually rely on the specification
of tuning parameters (in these cases threshold constants that
determine the extreme region), leading to suboptimal
solutions attained by means of intensive optimization routines. 

Dependence in extreme values of random variables  is assuming a
central role in risk management, and financial and insurance processes
offer, in this regard, an important source of examples. The high level
of integration in modern economies, the progress of information
technology, and the growing number of agents at play in financial
markets, are just some of the factors that seem to have contributed to
increased
dependence in the extremes of financial variables, as a growing number
of empirical studies are showing. \cite{Ramchand_1998},
\cite{Longin_2001}, and \cite{Campbell_2002} studied extreme
correlation in international equity markets. \cite{Poon_2004}
investigated the dependence in  both extreme losses and extreme gains between the S\&P
500 index and the various indices representing specific countries'
stock markets, finding that the former type of dependence is greater than the latter.   \cite{Hilal_2011} use the analysis of extremal dependency between the S\&P 500 index and VIX futures contracts in order to better hedge the risk of a long portfolio on the index. 
In some cases dependence among extremes  seems to be determined by the
structure of the phenomenon itself, as occurs for example between the daily
volatility of financial returns and the number of transactions
\cite{Rossi_2013}; or between extremes in returns and trading volumes
in Asian stock markets \cite{Ning_2009}.
In risk management, tail dependence is also strongly connected with
the concept of risk dependence \cite{Cherubini_2004, Embrechts_2002, Hilal_2011}.

In this work we propose new TDCs to quantify the degree of lower and
upper tail dependence between two random variables. The idea stems
from global dependence measures defined through surface integrals of the
copula function, suitably normalized with respect to the independence
case, and scaled by the difference between the
Fr\'echet-Hoeffding upper and lower bounds. These global measures are
developed in Section \ref{diffgeom}. The proposed TDCs,  introduced in Section \ref{diffgeom_2}, do
not rely on particular paths chosen in the copula domain, nor on
predefined regions of the support. They are able to detect positive
dependence structures in terms of joint probabilities greater than the
independence case, and in the joint tails for any asymptotic behaviour
of the associated marginals. They can also  provide information on the strength of
tail dependence, both in the case of asymptotic dependence, as well as
in the case of asymptotic independence.

The rest of the paper is organized as follows. A simulation study is
conducted in Section~\ref{sim_Analysis} in order to assess the performance of maximum
likelihood estimation in parametric modeling of the proposed TDCs,
while Section  \ref{CaseStudy} undertakes a  case study to illustrate
their usage in an empirical financial setting  investigating tail
dependence structures between some very visible stock indices. The
development of the global measures in Section \ref{diffgeom}
requires familiarity with fundamental concepts and results from copula theory and associated measures
of dependence and of concordance, and thus these are distilled into a
summary in Appendix \ref{copulatheory}. Proofs of more difficult theorems are
relegated to Appendix~\ref{AppendixB}, while a description of several copulas used 
throughout the paper is provided in Appendix~\ref{AppendixC}.

\section{Global Dependence}
\label{diffgeom}

Our proposed measure of tail dependence begins with a
global formulation aimed at the entire distribution, in analogy
with other global measures such as the correlation coefficient, Kendall's tau, and Spearman's rho.  Let $C(u,v)$ be the unique bivariate copula associated with the joint
distribution function $H(x,y)$ of the continuous bivariate random vector
$(X,Y)$, whose marginal cumulative distribution functions are $F(x)$
and $G(y)$, respectively. The material in this section assumes
familiarity with basic copula theory and associated measures of
dependence, a summary of which is provided in
Appendix~\ref{copulatheory}. Also, proofs of theorems not given
or alluded to in the text may be found in Appendix~\ref{AppendixB}.
\begin{definition}
For the  subsets of the unit square in $\mathbb{R}^2$
\[
  D_{v} = \left\{ (u,v): u \in [0,1], v \in (0,1] \right\},
  \qquad\text{and}\qquad D_{u} = \left\{ (u,v): u \in (0,1], v \in
    [0,1] \right\},
\]
let the functions $\Psi^{(l)}_{{X|Y}}: D_{v} \rightarrow [0,1]$ and
$\Psi^{(l)}_{Y|X}: D_{u} \rightarrow [0,1]$ be defined as
\begin{equation}\label{eq:CondCopula}
\Psi^{(l)}_{X|Y}(u,v):= \frac{ C(u,v) }{ v }, \qquad\text{and}\qquad \Psi^{(l)}_{Y|X}(u,v):= \frac{ C(u,v) }{ u },
\end{equation}
which represent the \emph{lower} cumulative conditional probabilities $P(X \le x |  Y \le y) $
and $P(Y \le y | X \le x) $, respectively.
\end{definition}

One way to describe the functions in \eqref{eq:CondCopula} is that they represent
conditionals in the direction of the SW portion of the unit square
(hence \emph{lower}). Based on the Fr\'echet-Hoeffding bounds $W$ and $M$ on any given copula
$C$ (see Theorem~\ref{th:FBbounds}), we can apply the same idea to
define the conditional Fr\'echet-Hoeffding boundary functions
\begin{align}
& \mathbb{W}_{X|Y}=\frac{W(u,v)}{v}, && \mathbb{W}_{Y|X}= \frac{W(u,v)}{u},
& \mathbb{M}_{X|Y}=\frac{M(u,v)}{v}, && \mathbb{M}_{Y|X}= \frac{W(u,v)}{u}.
\end{align}
Not surprisingly, the
unconditional Fr\'echet-Hoeffding bounds also apply in this conditional
probability framework.
\begin{theorem} \label{th:ConditionalFHbound}
For every copula $C(u,v)$ and every $(u,v)$ in $[0,1]\times(0,1]$, we
have
\begin{equation}
\label{eq:ConditionalFHbound_v}
\mathbb{W}_{X|Y}(u,v) \le \Psi^{(l)}_{X|Y}(u,v) \le \mathbb{M}_{X|Y}(u,v),
\end{equation}
and
\begin{equation}
\label{eq:ConditionalFHbound_u}
\mathbb{W}_{Y|X}(u,v) \le \Psi^{(l)}_{Y|X}(u,v) \le \mathbb{M}_{Y|X}(u,v).
\end{equation}
\end{theorem}
\begin{proof}
Consider $\Psi^{(l)}_{X|Y}$, then, given that $v$ can take values in $(0,1]$ and that copula functions are non-decreasing in each argument,
the two inequalities follow directly from Fr\'echet-Hoeffding
bounds. Similar reasoning applies to $\Psi^{(l)}_{Y|X}$.
\end{proof}

Note that if $C$ is a symmetric copula, then $\Psi^{(l)}_{X|Y}$ and $\Psi^{(l)}_{Y|X}$ are symmetric about the line $v=u$.
We now consider some properties of the functions $\Psi^{(l)}_{X|Y}(u,v)$. Similar
results follow straightforwardly for  $\Psi^{(l)}_{Y|X}(u,v)$. 
We start by studying the behaviour of function $\Psi^{(l)}_{X|Y}$ on the boundary of its domain: $\Psi^{(l)}_{X|Y}(u,1) = C(u,1) = u$, $\Psi^{(l)}_{X|Y}(1,v) = C(1,v) =v/v = 1$,
and $\Psi^{(l)}_{X|Y}(0,v) = 0/v =0$.
Now, $\Psi^{(l)}_{X|Y}(u,v)$ is not defined when $v=0$, but we can study the limit
$\lim_{v \to 0} \Psi^{(l)}_{X|Y}(u_{0},v)$ for some choice of $u_{0} \in [0,1]$.
Consider for instance the  functions $\mathbb{W}_{X|Y}$ and $\mathbb{M}_{X|Y}$. We can compute for any $u_{0} \in [0,1]$
\begin{align}
&\lim_{v \to 0 } \mathbb{W}_{X|Y}(u_{0},v) && =\lim_{v \to 0 } \frac{1}{ 2v }  \left( \ |u_{0}+v-1| +u_{0}+v-1 \right) =0, \\
&\lim_{v \to 0 } \mathbb{M}_{X|Y}(u_{0},v) &&=\lim_{v \to 0 } \frac{1}{ 2v }    \left( \ u_{0} + v - |u_{0}-v| \right) = 1. 
\end{align}
Now, given the above limits and the  two inequalities  in (\ref{eq:ConditionalFHbound_v}), we can state that  
\begin{equation} \label{eq:limit_bounds}
   0 \le \lim_{v \to 0} \Psi^{(l)}_{X|Y}(u_{0},v) \le 1, \qquad \text{for all }u_{0} \in [0,1]. 
\end{equation}
In other words,  the boundary $\partial \Omega$ of a surface like $\Psi^{(l)}_{X|Y}$ can be divided into two parts:
  $\partial \Omega_{1}$ and  $\partial \Omega_{2}$. The first part  is
  common to any surface 
  and can be defined as $\partial \Omega_{1}    =  D_{1} \cup D_{2} \cup D_{3}$, where
  $D_{1} = \left\{ (u,v,z): u=0 , v \in [0,1], z=0 \right\}$,
  $ D_{2} = \left\{ (u,v,z): u \in [0,1] ,  v = 1, z=u \right\} $, and  $ D_{3} = \left\{ (u,v,z): u =1 ,  v  \in [0,1], z=1 \right\}. $
The second part  will depend instead on the particular copula $C$
under consideration, and it  is bounded between 
$ \left\{ (u,v,z): u \in [0,1] ,  v = 0, z=0 \right\}$ and $\left\{ (u,v,z): u \in [0,1] ,  v = 1, z=0 \right\} $.  
Noting Theorem~\ref{th:C1deriv}, we can also state that $\partial \Omega_{2}$ will be a non-decreasing  
curve  from point $(0,0,0)$ to point $(1,0,1)$.
Since the $\Psi$ functions are, in general, not grounded, they are
also not copulas.
The partial derivatives for $\Psi^{(l)}_{X|Y}$  are:
\[
    \frac{\partial \Psi^{(l)}_{X|Y}(u,v)}{\partial u} := 
    \frac{1}{v} \frac{\partial C(u,v) }{\partial u},\qquad\text{and}\qquad
    \frac{\partial \Psi^{(l)}_{X|Y}(u,v)}{\partial v} := 
    \frac{1}{v} \frac{\partial C(u,v) }{\partial v}- \frac{C(u,v)}{v^2},
\]
where
\[
    P\left[ V \le v  \ | \ U =u \right] =  \frac{\partial C(u,v) }{\partial u},
    \qquad\text{and}\qquad P\left[ U \le u  \ | \ V =v \right] =  \frac{\partial C(u,v) }{\partial v}.
\]

The functions $\mathbb{M}$ and $\mathbb{W}$ represent the two extreme
situations of complete positive and negative dependence, where one
random variable is perfectly predictable from the other. In the
positive dependence situation of \emph{co-monotonicity}, one is a monotone increasing transformation
of the other, while for \emph{counter-monotonicity} the two random
variables are related by a monotone decreasing transformation. 
The independence case of no functional relationship is instead represented by 
\[
\mathbb{I}_{X|Y}:=\frac{\Pi(u,v)}{v}=u,\qquad\text{and}\qquad \mathbb{I}_{Y|X}:=\frac{\Pi(u,v)}{u}=v. 
\]
Indeed, given a point $\left( u_{0} , v_{0}\right)$ in $D_{v}$,
if $\Psi^{(l)}_{X|Y}(u_{0},v_{0}) > \mathbb{I}_{X|Y}(u_{0},v_{0})$, 
then
 $P(X \le x_{0} \text{ }|\text{ } Y \le y_{0}) > P(X \le x_{0} )$,
 and if 
$\Psi^{(l)}_{X|Y}(u_{0},v_{0}) < \mathbb{I}_{X|Y}(u_{0},v_{0})$,
we have the opposite inequality  
$P(X \le x_{0} \text{ }|\text{ } Y \le y_{0}) <P(X \le x_{0} ).$ 
Now, if we have instead $\Psi^{(l)}_{X|Y}=\mathbb{I}_{X|Y}$, 
then 
$\Psi^{(l)}_{_{Y|X}}=\mathbb{I}_{Y|X}$,  $P(X \le x_{0} \text{ }|\text{ } Y \le y_{0}) = P(X \le x_{0})$, 
and $P(Y \le y_{0} \text{ }|\text{ } X \le x_{0}) = P(Y \le y_{0})$.\\

\begin{figure}
	\centering	
        \subfloat[  ]{\includegraphics[scale= 0.5]{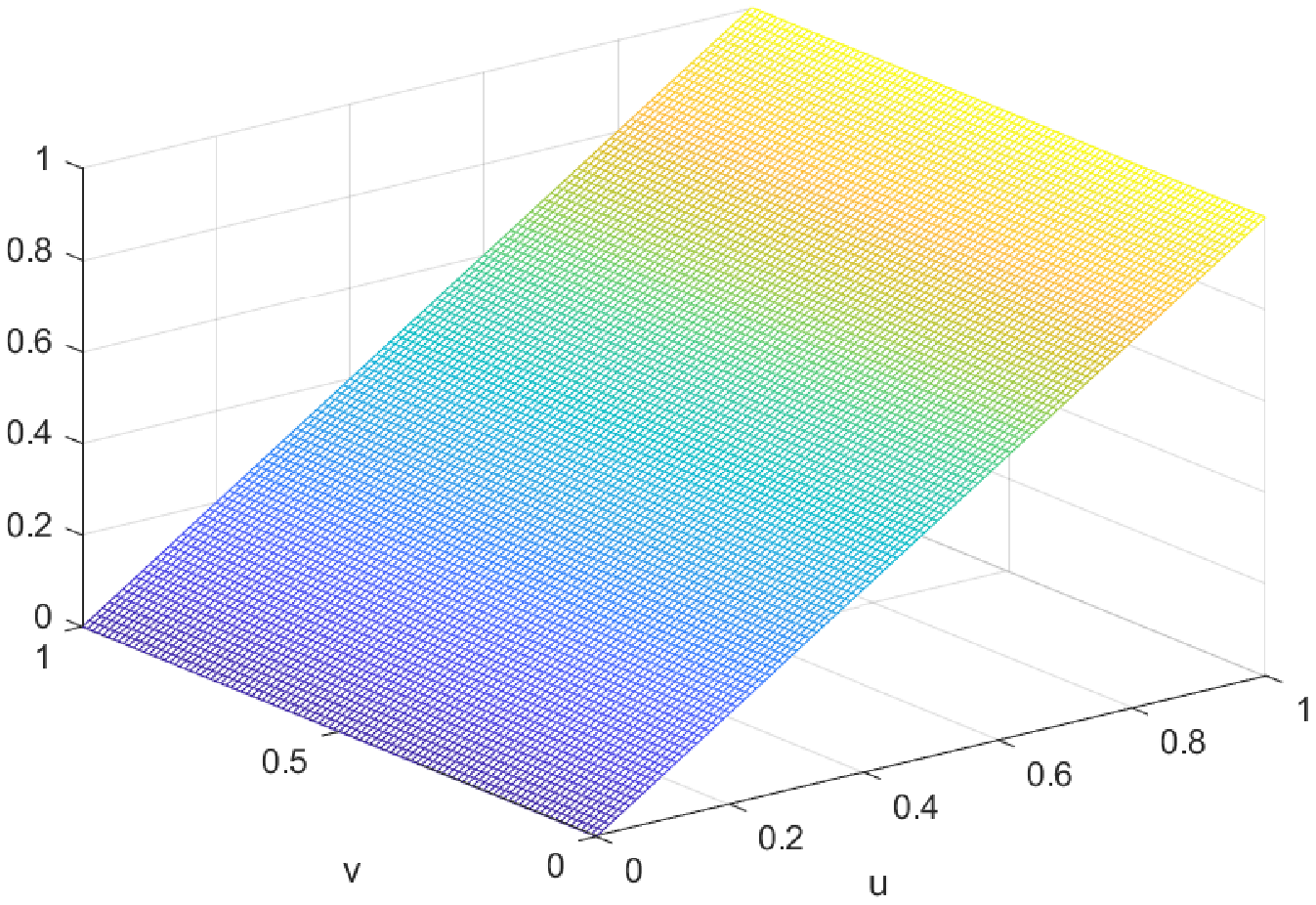}}
	\subfloat[  ]{\includegraphics[scale= 0.43]{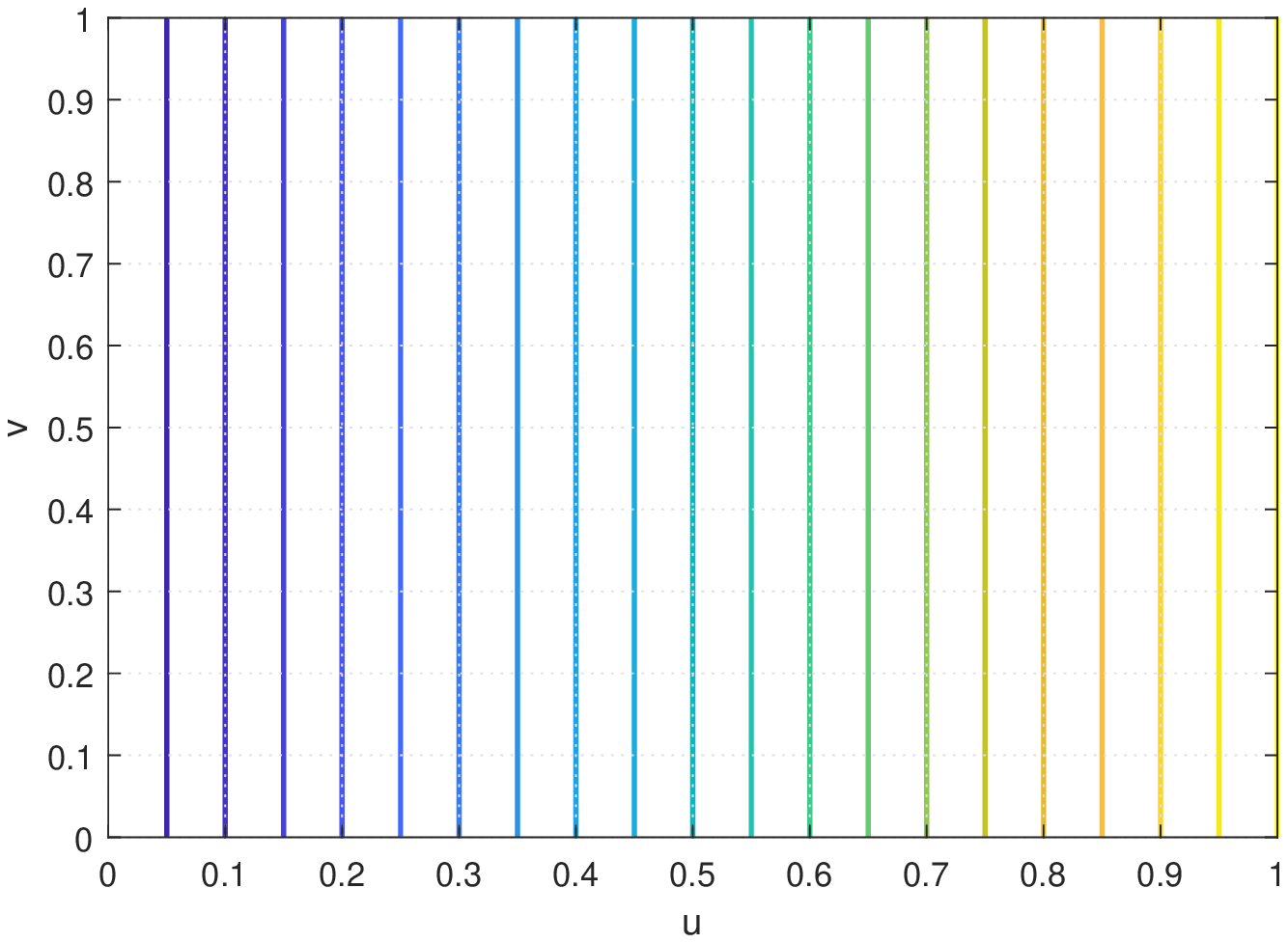}}\\
	\subfloat[  ]{  \includegraphics[scale= 0.5]{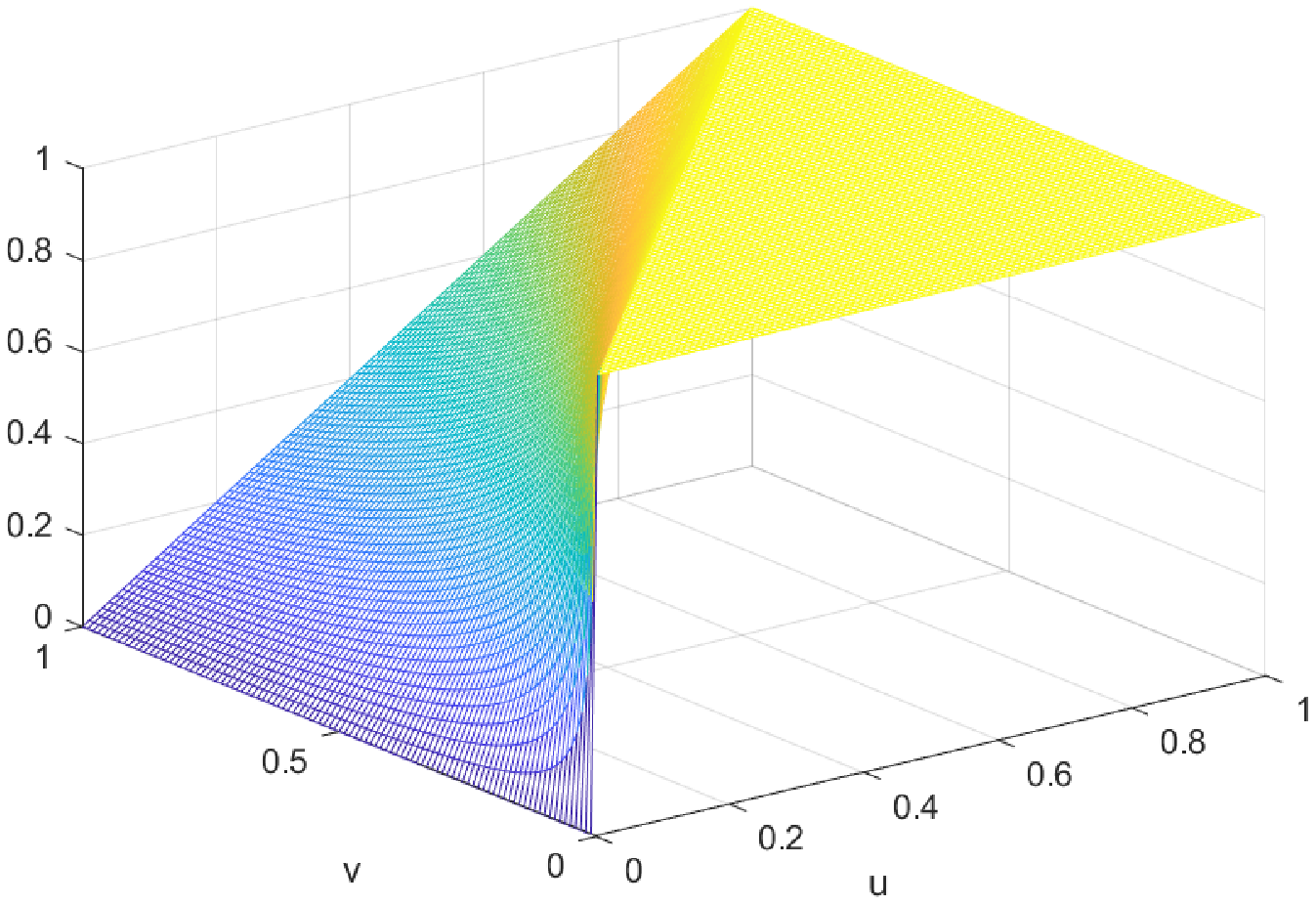} }
         \subfloat[  ]{  \includegraphics[scale= 0.43]{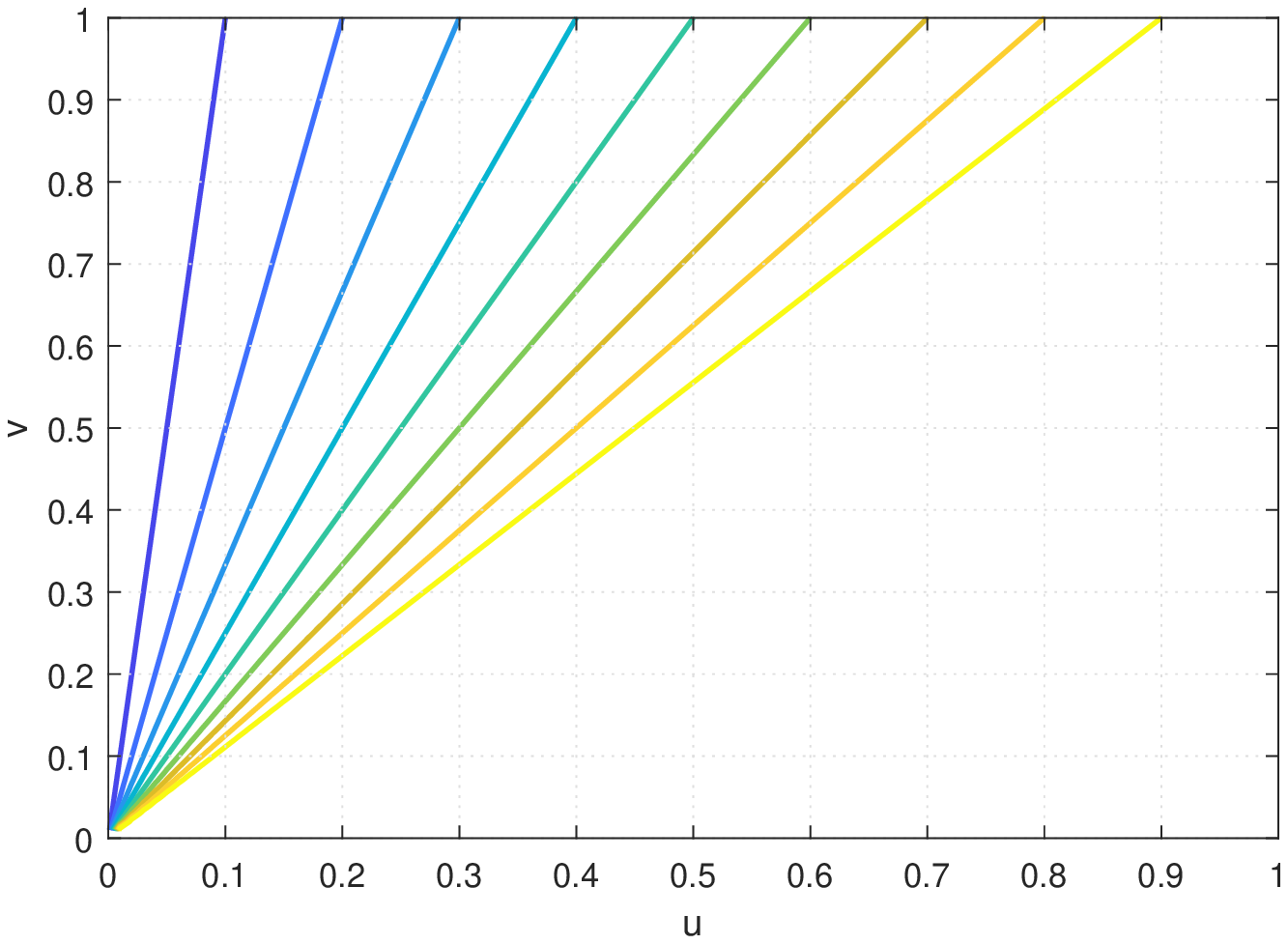} }\\	
	\subfloat[  ]{\includegraphics[scale= 0.38]{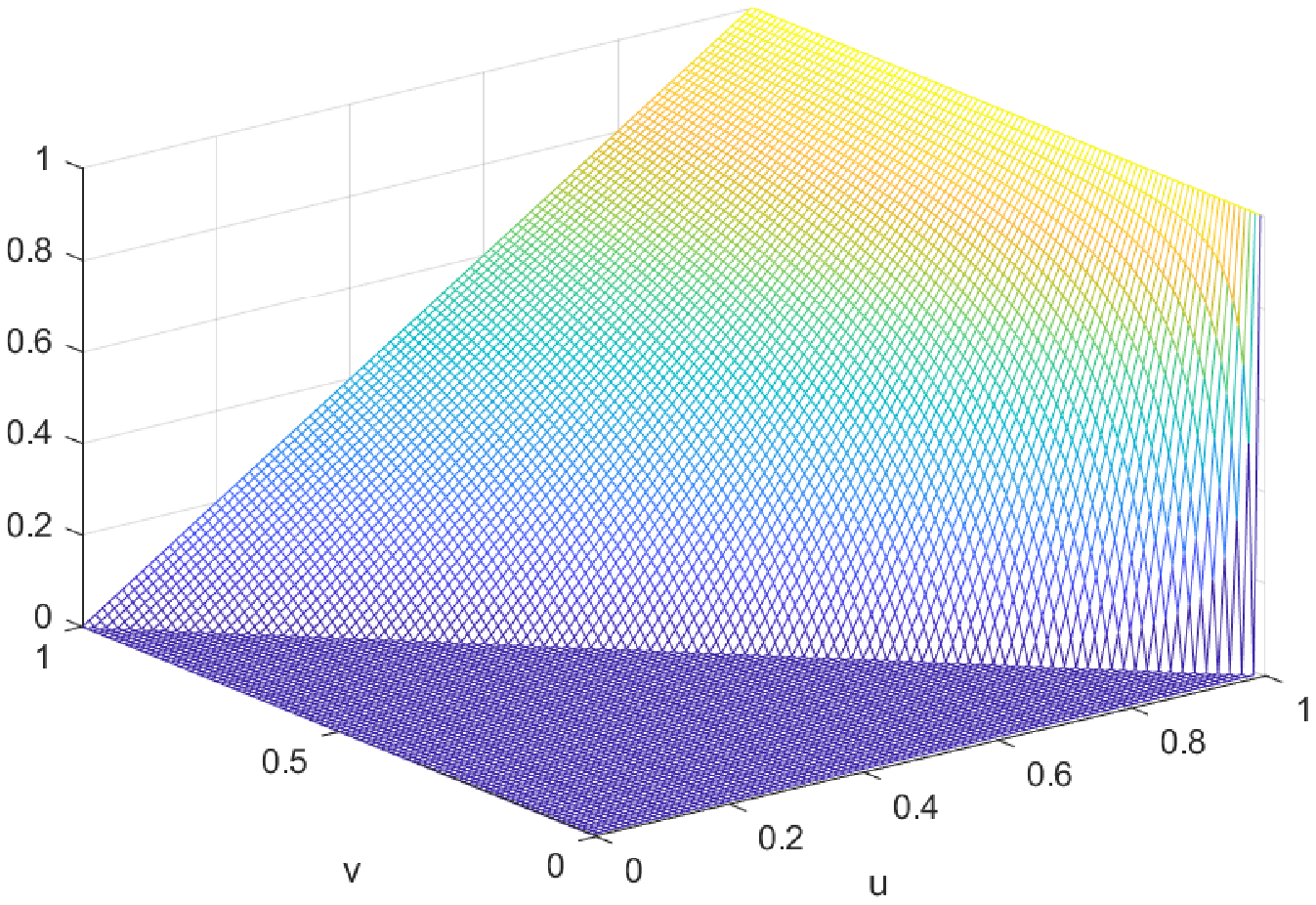}}
	\subfloat[  ]{\includegraphics[scale= 0.31]{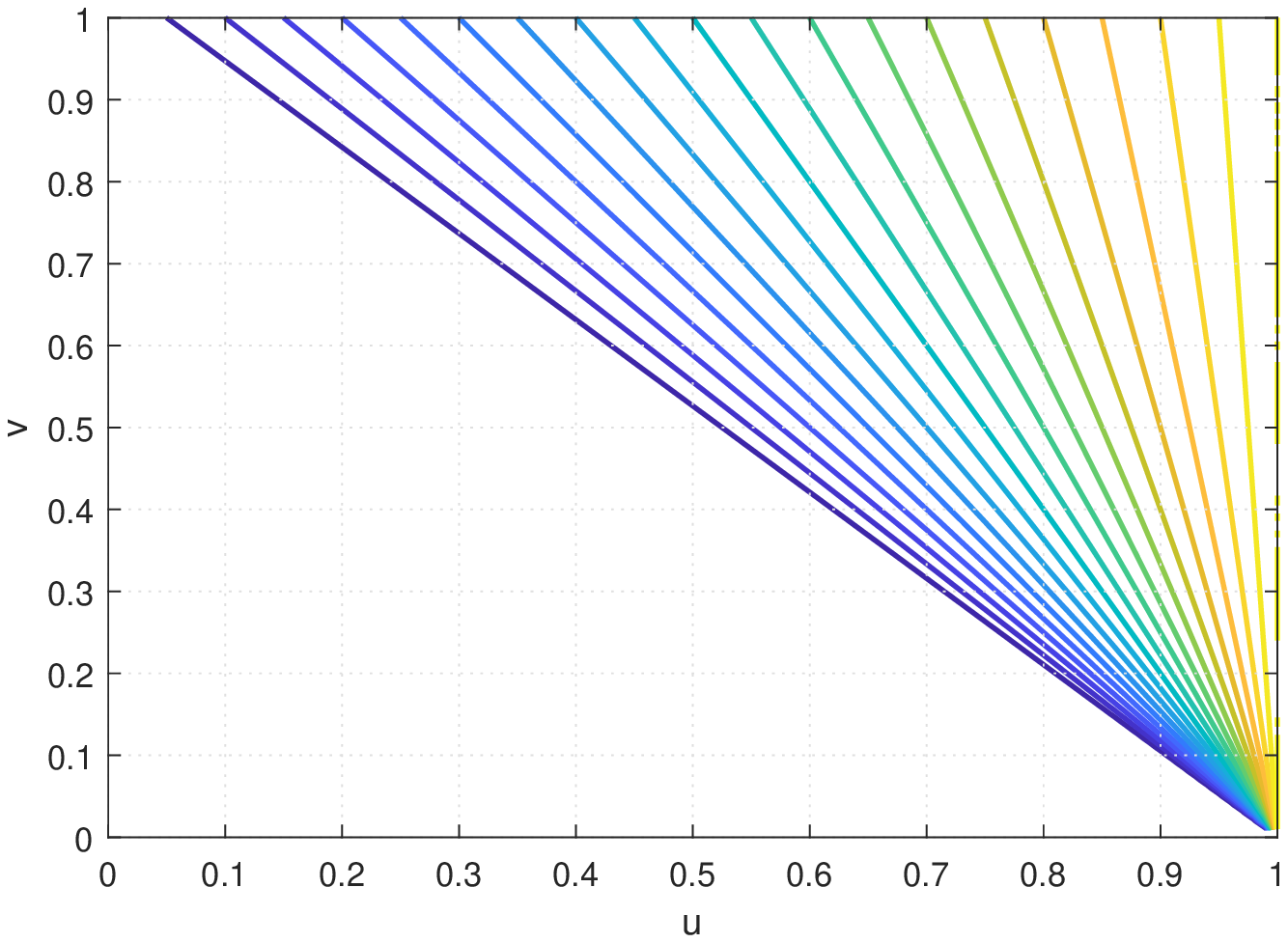}}		
	\caption{Lower unit square cumulative conditional probability surfaces (left panels) and their level curves (right panels) in copula space corresponding to the cases of independence ($\mathbb{I}_{X|Y}$, top panels), co-monotonicity ($\mathbb{M}_{X|Y}$, middle panels), and counter-monotonicity ($\mathbb{W}_{X|Y}$, bottom panels). }
	\label{fig:Surf_LIU}
\end{figure}

\noindent Functions  $\mathbb{I}_{X|Y}$, $\mathbb{M}_{X|Y}$, and $\mathbb{W}_{X|Y}$ are displayed in Figure \ref{fig:Surf_LIU}, along with their level curves.
From a geometric point of view, the function $\mathbb{I}(u,v)$ represents the cylinder with generatrix line $z=u$ and directrix the $v$-axis enclosed in the unit cube, 
which is simply a plane connecting the lower-left and the upper-right
sides of the unit cube. Noting  that the given plane is a \textit{minimal surface} inside
 the unit cube, we can state the following theorem.

 \begin{theorem} \label{th:Asurf_disequality}
   Let $C(u,v)$ be a bivariate copula and 
       $A( \Psi^{(l)}_{X|Y}):=\int_{\Psi^{(l)}_{X|Y}} 1 \text{ }dS$ denote  the integral of $1$ over the surface 
       $\Psi^{(l)}_{X|Y}(u,v):=C(u,v)/v$, i.e, the surface area  of
       $\Psi^{(l)}_{X|Y}$, then we have that
    \begin{equation} \label{eq:Asurf_disequality}
           \sqrt{2}= A(\mathbb{I}_{X|Y}) \leq A(\Psi^{(l)}_{X|Y}) \leq A(\mathbb{M}_{X|Y}) 
           =A(\mathbb{W}_{X|Y})=1.708.
     \end{equation}
\end{theorem}
While it is straightforward to analytically compute the lower bound of
$\sqrt{2}$, the upper bound of $1.708$ was obtained numerically (although
it is possible to solve one of the two integrals analytically so that the numerical integration is in just one dimension). 

Theorem~\ref{th:Asurf_disequality} suggests that the surface area of
$\Psi^{(l)}_{X|Y}$ could be used as a measure of dependence. Indeed, one can
define an index by normalizing the surface area $A(\Psi^{(l)}_{X|Y})$ over the interval $[0,1]$. 
Of course, the same reasoning applies to $\Psi^{(l)}_{Y|X}$; furthermore, it is easy to show that 
$A(\mathbb{W}_{Y|X})=A(\mathbb{W}_{X|Y})$,
$A(\mathbb{M}_{Y|X})=A(\mathbb{M}_{X|Y})$ and
$A(\mathbb{I}_{Y|X})=A(\mathbb{I}_{X|Y})$,  so that we can remove subscripts.
This therefore motivates the definition of $\delta^{(l)}_{X|Y}$ given
below. Note that (obvious) analogous results hold for $\delta^{(l)}_{Y|X}$ defined through
$\Psi^{(l)}_{Y|X}$, here and in the ensuing discussion, and we therefore
omit any further allusions to it.
\begin{definition}\label{def:delta-dependence}
Let $X$ and $Y$ be two continuous random variables with joint
distribution function $H$ and copula $C$. Let $\Psi^{(l)}_{X|Y}$ and
$\Psi^{(l)}_{Y|X}$ be as defined in \eqref{eq:CondCopula}. We define the
lower (unit square cumulative conditional probability) measure of dependence
$\delta^{(l)}_{X|Y}$ as
\begin{equation}\label{eq:delta-lower}
\delta^{(l)}_{X|Y} := \frac{ A(\Psi^{(l)}_{X|Y}) - A(\mathbb{I} ) }{
  A(\mathbb{W}) - A(\mathbb{I} ) }= \frac{ A(\Psi^{(l)}_{X|Y}) -
  A(\mathbb{I} ) }{ A(\mathbb{M}) - A(\mathbb{I}) }.
\end{equation}
\end{definition}

In order to more fully understand  $\delta^{(l)}_{X|Y}$,
consider the vector field
$\nabla{\Psi^{(l)}_{X|Y}}$ that acts as a weight with respect to the infinitesimal area element
$dudv$ in the surface integral 
\[
A(\Psi^{(l)}_{X|Y}) = \int_{\Psi^{(l)}_{X|Y}} \sqrt{1+ ||\nabla{\Psi^{(l)}_{X|Y}(u,v)}||^{2}}\;dudv.
\]
In the independence case we have that
$\partial\Psi^{(l)}_{X|Y}(u,v)/\partial v=0$ and $\partial
\Psi^{(l)}_{X|Y}(u,v)/\partial u=1$ for all $(x,y) \in D_{v}$. 
With respect to the upper  
bound of $\mathbb{M}_{X|Y}$,
it is easy to see that 
$\partial \mathbb{M}_{X|Y}(u,v)/\partial u$
is everywhere non-negative, assuming a maximum value as $u,v \to 0$, whereas
$\partial\mathbb{M}_{X|Y}(u,v)/\partial v$ is always non-positive,
assuming a minimum value as $u,v \to 0$. 
Similarly, observe that  
$\partial\mathbb{W}_{X|Y}(u,v)/\partial u$ and $\partial\mathbb{W}_{X|Y}(u,v)/\partial v$
are everywhere non-negative, taking on maximum values as  $u \to 1$ and $v \to 0$.
These considerations imply that as the function $\Psi^{(l)}_{X|Y}$ converges
toward $\mathbb{M}_{X|Y}$ or $\mathbb{W}_{X|Y}$, the value of
$||\nabla{\Psi^{(l)}_{X|Y}}||$ increases, and does so at an increasingly
faster rate as it approaches the point $(0,0)$ in the case of positive
dependence, and the point $(0,1)$ in the case of negative
dependence. In other words, $\delta^{(l)}_{X|Y}$ assigns
increasing weight to those points $(u,v)$ that depart from the
independence case and that are either in the lower-left or lower-right portion of the domain $D_{v}$.

The next result states that $\delta^{(l)}_{X|Y}$ satisfies the properties of
a \emph{dependence measure}, as stated in Definition
\ref{def:Dependence}, except for property 3 which requires symmetry
with respect to conditioning; $X|Y$ giving the same
measure as $Y|X$. 
\begin{theorem}
\label{th:DepMeasure}
The measure $\delta^{(l)}_{X|Y}$ in Definition
\ref{def:delta-dependence} satisfies all the properties of a measure
of dependence according to Definition \ref{def:Dependence},  except for property 3.
\end{theorem}

We argue that this is a desirable feature of the measures $\delta^{(l)}_{X|Y}$
and $\delta^{(l)}_{Y|X}$, since it more completely accounts for asymmetric
dependence structures. This follows from the fact that in general $C(u,v)\neq
C(v,u)$, thus implying different dependence relationships
with respect to the direction of conditioning. This is clear when one
realizes that \textit{dependence} in probability theory is defined as
the absence  of  \textit{independence};
 i.e., we have dependence every time we do not have independence. But independence, after all, is  defined through conditional probabilities, 
 which implies that  deviation from independence can arise 
 differentially with respect to  the direction of conditioning. 
 In summary, we argue that property  3 in
 Definition~\ref{def:Dependence} should be excluded from the axiomatic
 list of dependence measures.

The fact that the normalization in Definition
\ref{def:delta-dependence} is insensitive to whether co-monotonicity
or counter-monotonicity is used, means that the ``sign'' of dependence
is lost. In order to devise a measure that differentiates
between positive and negative dependence  structures, what is defined in the
literature as a measure of \emph{concordance}, 
a simple and  intuitive approach is to consider the surface integral
\[
    S_{2}(\Psi^{(l)}_{X|Y}) := \int_{ \Psi^{(l)} } \left[ \  \Psi^{(l)}_{X|Y}(u,v) - \mathbb{I}_{X|Y} (u,v) \  \right] \text{ }dS.
\]
For any given point $\left(u_{0},v_{0}\right)$, this integral spans the volume of the infinitesimal
parallelepiped with base area $\left(1+ ||\nabla{\Psi^{(l)}_{X|Y}}||^{2}\right)^{1/2} \ du_{0} \ dv_{0}$ and  height $ |\Psi^{(l)}_{X|Y}(u,v) - \mathbb{I}_{v} (u,v)|$. 
Suitable normalization of $S_{2}$ values over the interval $[-1,1]$,
leads to the definition of the \emph{lower} (unit square cumulative conditional probability) concordance index
\begin{equation}\label{eq:kappa-lower}
\kappa^{(l)}_{X|Y} := \frac{ 2\left[ S_{2}(\Psi^{(l)}_{X|Y}) -S_{2}(\mathbb{W}) \  \right] }{ S_{2}(\mathbb{M}) - S_{2}(\mathbb{W}) } -1,
\end{equation} 
Using similar reasoning as for Theorem \ref{th:DepMeasure}, it is
straightforward to show that $\kappa^{(l)}_{X|Y}$ satisfies all the
properties of a measure of concordance listed in
Definition~\ref{def:Concordance}, except for property 3. Kendall's tau
$(\tau)$ and Spearman's rho $(\rho)$ are perhaps the best known  measures of concordance satisfying all 7 properties.

The definition of the measures $\delta^{(l)}_{X|Y}$ and
$\kappa^{(l)}_{X|Y}$ relied on the cumulative
conditional probability $P(X\leq x|Y\leq y)$ in order to quantify departures from
independence that occur on the \emph{lower} triangular half of the
unit square bounded by the line $u+v=1$. Similarly, and with
special relevance to the introduction of tail dependence in
the next section, we can apply the ideas developed so far to devise
analogous measures that focus on the \emph{upper} triangular half, based on the conditional probability $P(X>
x|Y> y)$. The development parallels the earlier, starting with the
definition of corresponding subsets of the unit square
\[
D_{1-v} = \left\{ (u,v): u \in [0,1], v \in [0,1) \right\}, \qquad\text{and}\qquad D_{1-u} = \left\{ (u,v): u \in [0,1), v \in [0,1] \right\},
\]
leading to the introduction of functions $\Psi^{(u)}_{{X|Y}}: D_{1-v} \rightarrow [0,1]$ and $\Psi^{(u)}_{Y|X}: D_{1-u} \rightarrow [0,1]$,
\begin{align} \label{eq:CondCopulaU}
    &\Psi^{(u)}_{X|Y}(u,v):= \frac{ 1-u-v+C(u,v) }{ 1-v }, &&\Psi^{(u)}_{Y|X}(u,v):= \frac{ 1-u-v+C(u,v) }{ 1-u },
\end{align}
which represent $P(X > x |  Y > y) $ and $P(Y > y | X > x) $,
respectively. It is straightforward to show that
$(1-u-v+\mathbb{W})/(1-v)$ and $(1-u-v+\mathbb{W})/(1-u)$ can be
obtained by respectively rotating $\mathbb{W}/v$ and $\mathbb{W}/u$
about  the $z$-axis through $\pi$ radians.
This leads directly to the definition of the \emph{upper} triangular
half versions
of \eqref{eq:delta-lower} and \eqref{eq:kappa-lower},
\[
\delta^{(u)}_{X|Y} := \frac{ A(\Psi^{(u)}_{X|Y}) - A(\mathbb{I} ) }{
  A(\mathbb{W}) - A(\mathbb{I} ) }= \frac{ A(\Psi^{(u)}_{X|Y}) -
  A(\mathbb{I} ) }{ A(\mathbb{M}) - A(\mathbb{I}) },
\]
and
\[
\kappa^{(u)}_{X|Y} = \frac{ 2\left[ S_{2}(\Psi^{(u)}_{X|Y}) -S_{2}(\mathbb{W}) \  \right] }{ S_{2}(\mathbb{M}) - S_{2}(\mathbb{W}) } -1.
\] 

We illustrate the values of both lower and
upper versions of $\kappa_{X|Y}$ along with Kendall's tau and Spearman's rho, for a select
list of copulas  in Tables~\ref{tab:kappa_GH}
and~\ref{tab:kappa}. In Table~\ref{tab:kappa_GH}, the copulas used
are implied by the GH distributional parameters listed in
Table~\ref{tab:GH_Table}, which implicitly define the copula (hence
``implied''). These GH's  have a relatively high
number of parameters (approximately 10).  Functional forms for the
copulas of Table~\ref{tab:kappa} can be found in Appendix \ref{AppendixC}. Results are obtained with
numerical integration on a structured Cartesian  mesh with $10^{6}$
points (although in some cases it may actually be possible to obtain closed formulas). 

We observe that the $\kappa$ measures assign more weight when the
dependence, whether positive or negative, is in the extreme zones of the support.
 Compare for instance the value of  $\kappa^{(l)}$ for  the Gumbel copula with $\theta=4$ and the Clayton copula with $\theta=5$.
  These copulas have relatively close values for each of Kendall's
  $\tau$ and Spearman's $\rho$, but larger differences between their
  corresponding $\kappa^{(l)}$, with the value of 0.92 for the Clayton
  exceeding that of 0.84 for the Gumbel. The situation then reverses
  for $\kappa^{(u)}$, where the Clayton value of 0.73 is smaller than
  the 0.91 of the Gumbel. These results are consistent with  the fact
  that the Clayton and Gumbel copulas have, respectively, Sibuya upper and lower TDCs equal to
  zero.

\begin{figure}
	\centering	
	\subfloat[  ]{\includegraphics[scale= 0.4]{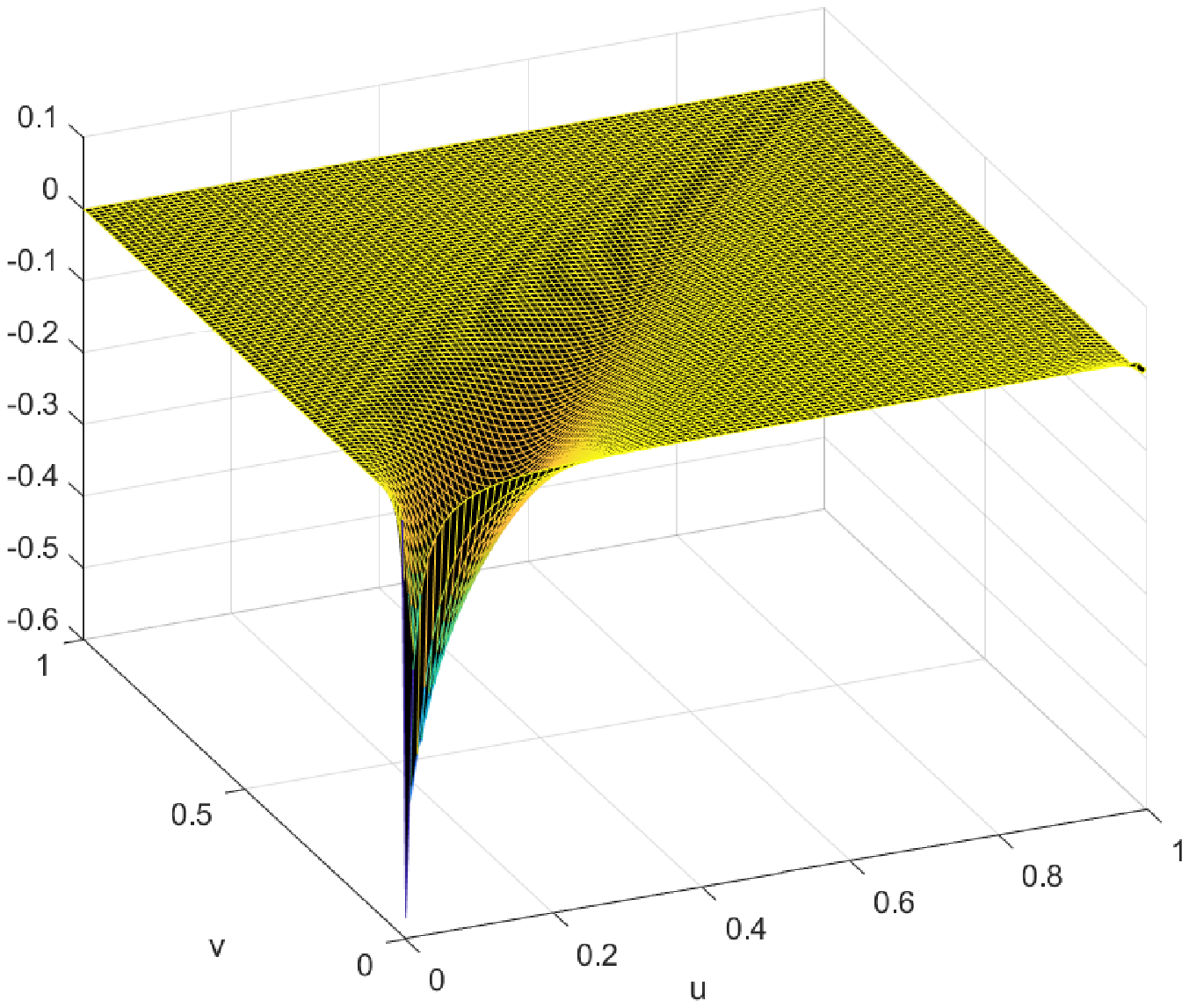}}
	\subfloat[  ]{\includegraphics[scale= 0.4]{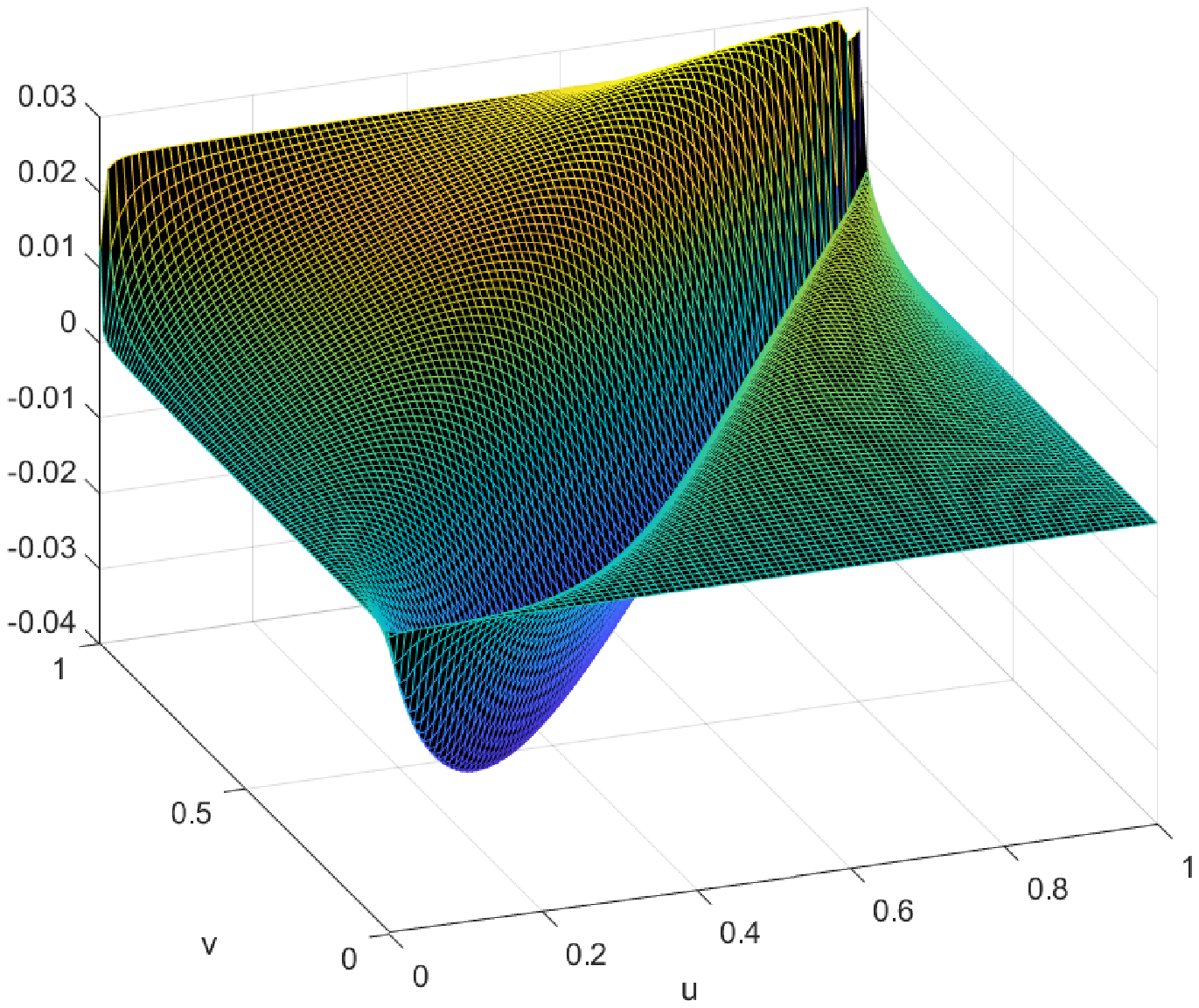}}
	\caption{Panel (a): difference in the $\Psi^{(l)}_{X|Y}$ surface
          between the Gumbel copula with parameter $\theta=4$ and 
	the $t$ copula with parameters $\nu=1$ and $\rho=0.95$. Panel (b):  the same difference, but with respect to the surface $\Psi^{(u)}_{X|Y}$. }
	\label{fig:psi_theta}
\end{figure}

As another example, compare  the Gumbel with $\theta=4$ and the
$t$ copula with $\nu=1$ and $ \rho=0.95$.  The Gumbel has lower
values of $\kappa^{(l)}$  because its $\Psi$ function is always
smaller. On the other hand, $\kappa^{(u)}$ values are very close since
neither of the two $\Psi^{(u)}$ surfaces dominates the other. We can
see this in Figure \ref{fig:psi_theta}, where panel (a) displays the
difference in the $\Psi^{(l)}_{X|Y}$ surfaces between Gumbel and $t$, whereas
panel (b) shows this  difference with respect to the $\Psi^{(u)}_{X|Y}$ surface.
Finally,  it is evident that in either the $\Psi^{(l)}$ or $\Psi^{(u)}$ case, there is generally no difference between
conditioning on $X|Y$ or $Y|X$. Some Marshall copulas and 
asymmetric GH distributions however, provide some examples to the contrary.

An alternative definition of $\delta^{(l)}_{X|Y}$ could be achieved through the surface integral 
\[
   S_{1}(\Psi^{(l)}_{X|Y})  := \int_{ \Psi^{(l)}_{X|Y} } \left| \Psi^{(l)}_{X|Y}(u,v) - \mathbb{I}_{X|Y}(u,v) \right| \text{ }dS,
\]
whence scaling by either boundary copula would lead to the measure
\[
  \bar{ \delta }^{(l)}_{X|Y}  :=  \frac{ S_{1}(\Psi^{(l)}_{X|Y})  }{
    S_{1}(\mathbb{W} )  }= \frac{ S_{1}(\Psi^{(l)}_{X|Y})  }{
    S_{1}(\mathbb{M})  }.
\]
This idea paves the way for the introduction of a TDC in the next section.

\section{Tail  Dependence}\label{diffgeom_2}

The previous section introduced  global measures that focused on
summarizing the dependence structure as revealed by surface integrals
of the lower and
upper  cumulative conditional probabilities in copula space.
In this section we adapt these measures to assess lower and
upper tail dependence in extreme regions of the copula domain, leading
to the definition of TDCs in similar vein to the weak
($\chi$) and strong ($\lambda$) discussed in the Introduction.

For a function $\mathbb{Q}(u,v)$
serving as a place-holder for any of the surfaces $\Psi$, $\mathbb{M}$, $\mathbb{W}$, or
$\Theta$ in the copula domain,  define the operator  $L(\cdot)$ as 
\[ 
L\left(\mathbb{Q}(u,v) \right):= -\min \left\{ \frac{\partial\mathbb{Q}(u,v)}{\partial v} \cdot \frac{\partial\mathbb{Q}(u,v)}{\partial u} \ , \ 0 \right\},
\]
and the corresponding surface integral
\begin{equation*}\label{eq:placeholder-surf-int}
\Gamma(\mathcal{Q};p) := \int_{ \mathcal{Q} }
\left[ L\left(\mathcal{Q} (u,v) \right)\right]^{p}\;dS.
\end{equation*}
With this notation, observe that $L(\Psi^{(l)}_{X|Y}(u,v))$ will assume greater
values  as the difference  $ \mathbb{M}_{X|Y}(u,v) - \Psi^{(l)}_{X|Y}(u,v)$
decreases. Furthermore, as $\Psi^{(l)}_{X|Y}$ converges toward
$\mathbb{M}_{X|Y}$, $L(\Psi^{(l)}_{X|Y}(u,v))$ increases faster as  $u$ and
$v$ approach zero. This ties in with the behavior of   
$\partial\mathbb{M}_{X|Y}(u,v)/\partial u$ as already noted in
the previous section, non-negative assuming a maximum value as $u,v \to 0$,
and that of  
$\partial\mathbb{M}_{X|Y}(u,v)/\partial v$
which is non-positive and is minimized as $u,v \to 0$. The same
analysis holds for $L(\Psi^{(l)}_{Y|X})$.

We can now introduce the \emph{lower} and \emph{upper} TDCs defined through the normalized
ratio of surface integrals,
\begin{equation} \label{eq:Lambda_tail_l}
\Lambda^{(l)}_{X|Y}(p) := \frac{  \Gamma( \Psi^{(l)}_{X|Y} ; p ) }{
  \Gamma(\mathbb{M}_{X|Y};p)  }, \qquad\text{and}\qquad \Lambda^{(u)}_{X|Y}(p) = \frac{\Gamma( \Psi^{(u)}_{X|Y};p)}{\Gamma(\mathbb{M}_{X|Y};p)}. 
\end{equation} 
Note that normalization by
$\Gamma(\mathbb{W}_{X|Y};p)$ would not result in a valid measure, as
this surface integral is zero. In order to do so, one would have to
modify the definition of $L(\cdot)$ by replacing $-\min$ with $\max$. Normalization by
$\Gamma(\mathbb{W}_{X|Y};p)$ would now result in a measure of
\emph{counter-tail} (opposite) dependence, i.e., $P(X<  x| Y>y )$ for small $x$ and large $y$.
As such, the choice of normalizing by $\Gamma(\mathbb{M}_{X|Y};p)$
provides a measure of \emph{co-tail} dependence (similarity), in the
spirit of the weak and strong TDCs alluded to in the Introduction.

\begin{figure}
	\centering		
	\includegraphics[scale=.7]{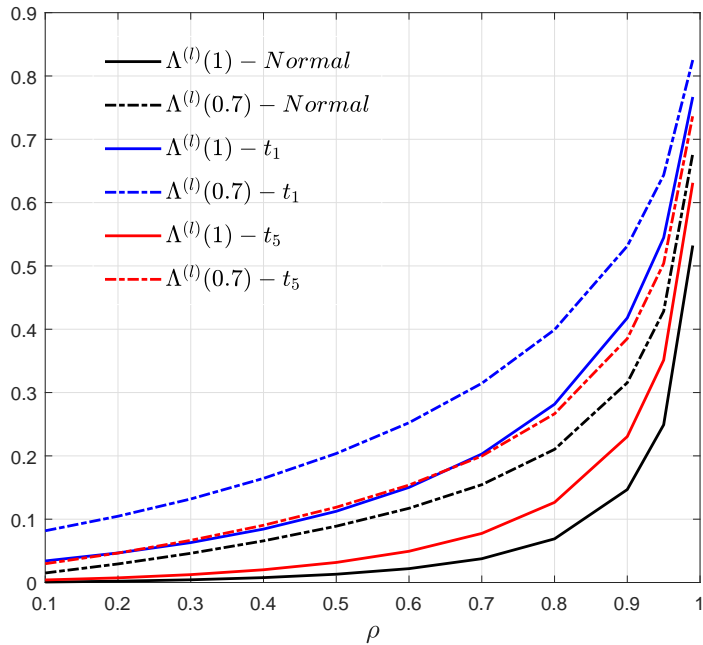}	
	\caption{Lower TDC $\Lambda^{(l)}(p)$ for the
          Gaussian and $t$ distributions as a function of the correlation
	parameter $\rho$.} 
	\label{fig:tail_Elliptic}
\end{figure}

The parameter $p$ can be used to increase, if greater than one, or
to decrease, if less than one, the focus on extreme sets of the joint
tail. In the case of $\Lambda^{(l)}$ this occurs when both $u$ and $v$
approach zero, whereas for $\Lambda^{(u)}$ it corresponds to $u$ and $v$
approaching one.
Indeed, in the former case given that the product of $\partial\Psi^{(l)}_{X|Y}(u,v)/\partial
v$ and $\partial\Psi^{(l)}_{X|Y}(u,v)/\partial u$ grows at an increasingly
faster rate as $\Psi^{(l)}_{X|Y}(u,v)$ converges to $\mathbb{M}_{X|Y}(u,v)$,
it will be convenient in some cases to select a value of $p<1$ in
order to re-scale $\Lambda^{(l)}$. 
 We provide an example in Figure \ref{fig:tail_Elliptic} which compares $\Lambda^{(l)}$
  for the Gaussian and $t$ distributions as a function of the
  correlation parameter $\rho$. Results for $p=1$ ($p=0.7$) are
  plotted with solid (dotted) lines.  Note that for  $p<1$,
  the fact that $\Lambda^{(l)}(p)$ assigns less weight to the extreme regions results
  in a more gradual variation and smaller range in its
  values. Similar observations hold for $\Lambda^{(u)}$.  

Table ~\ref{tab:Lambda_GH} displays the values of $\Lambda^{(l)}_{X|Y}(p)$,
$\Lambda^{(l)}_{Y|X}(p)$, $\Lambda^{(u)}_{X|Y}(p)$, and
$\Lambda^{(u)}_{Y|X}(p)$, for the GH copulas investigated in the global
dependence measures (Section~\ref{diffgeom}), and for the two values of $p=\{0.7,1\}$. The  last two columns in these tables
correspond to the strong lower and upper TDCs, $\lambda_{l}$ and $\lambda_{u}$, and
provide a meaningful comparison. Results are computed via numerical
integration on a uniform mesh with $10^{6}$ points.  Observe how all the chosen GH copulas
have positive lower and upper $\Lambda$ values, but only the case of
$GH_2$ exhibits asymptotic tail dependence ($\lambda_{l}=1=\lambda_{u}$).

Tables~\ref{tab:Lambda_1} and \ref{tab:Lambda_2} display the same
information, but for the non-GH copulas of Section~\ref{diffgeom}. For
readibility Table~\ref{tab:Lambda_1} (Table~\ref{tab:Lambda_2}) deals only with the case  
$p=1$ ($p=0.7$).
Note how the Gumbel (Clayton) copula has a lower (upper) strong TDC
of $\lambda=0$, but the $\Lambda^{(l)}$ ($\Lambda^{(u)}$) values
are increasingly different from zero as the copula parameter $\theta$
increases. The proposed $\Lambda$ TDCs also correctly identify
asymmetries with respect to the direction of conditioning, $X|Y$ or
$Y|X$. This is the case, for example, for the Marshall copula with
parameters $\{0.7,0.9\}$, or the $GH_{2}$ and $NIG_{4}$ copulas,
where the asymmetry is determined  by $\beta$.

\begin{figure}
	\centering	
	\subfloat[  ]{\includegraphics[scale= 0.37]{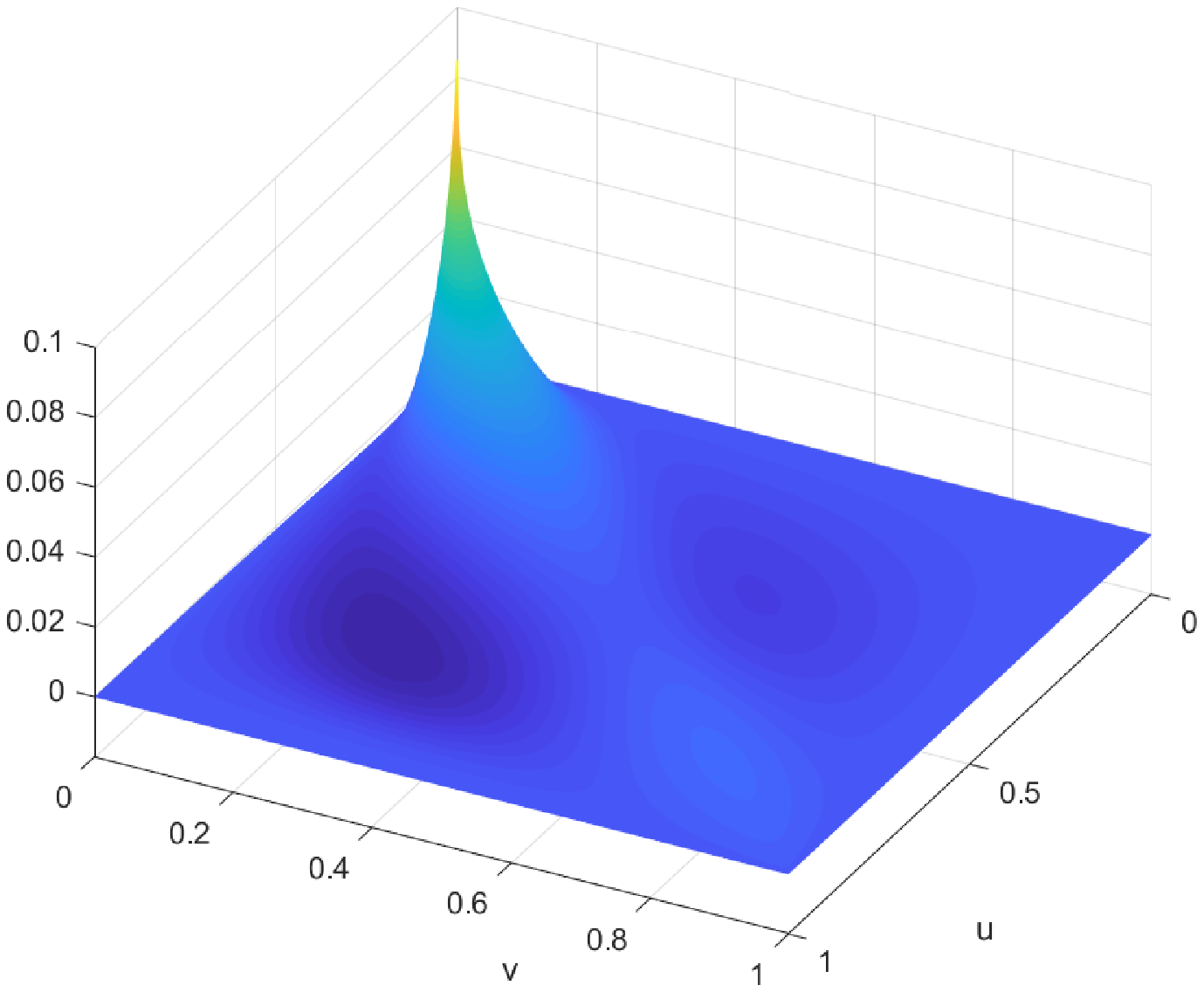}}
	\subfloat[  ]{\includegraphics[scale= 0.37]{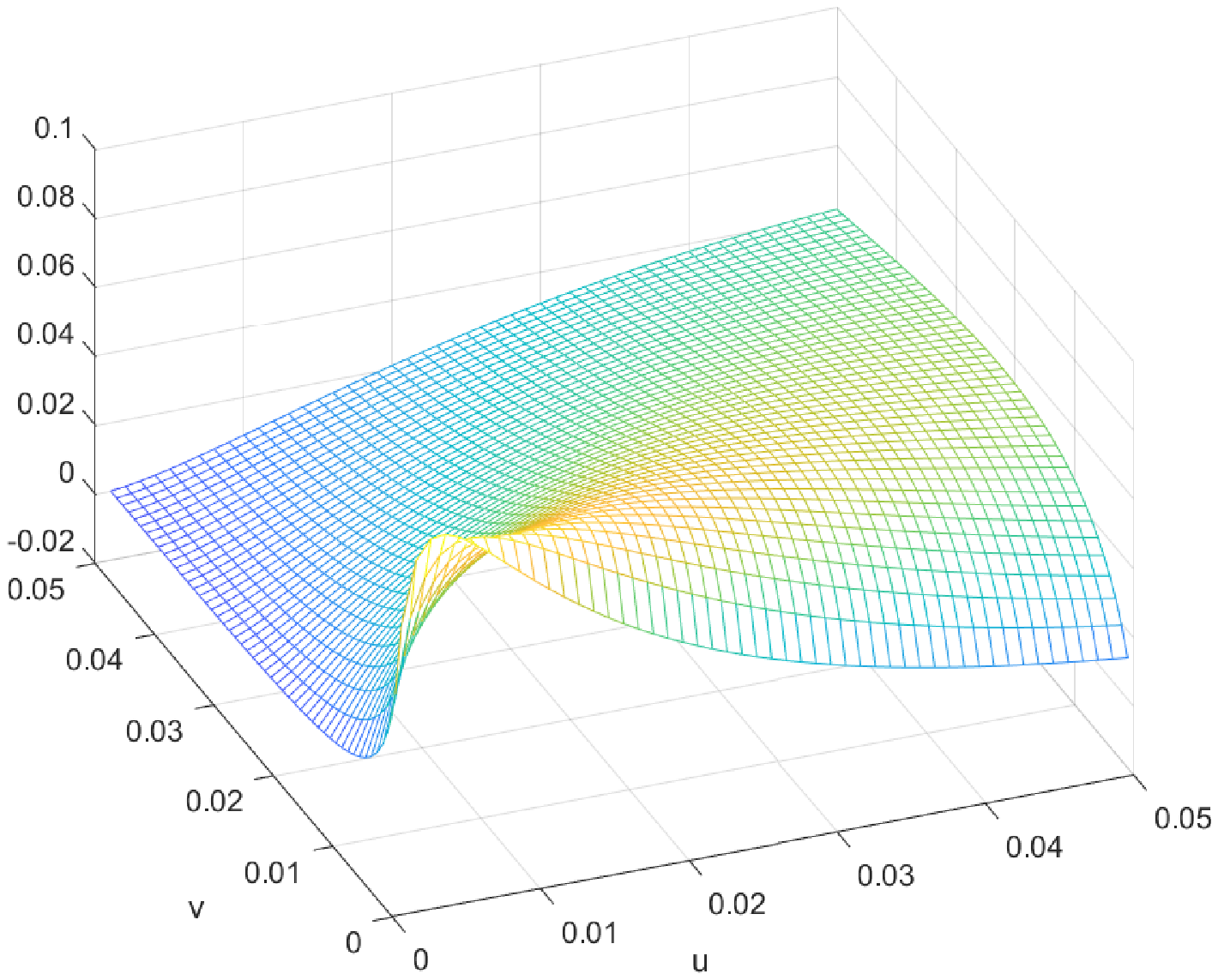}}\\
	\caption{Difference in the $\Psi^{(l)}_{X|Y}$ surfaces for a symmetric GH
distribution with parameters $\lambda=1.5$, $\delta=1$, $\mu=[0,0]$,
$\Delta_{12}=0.8$, and the two different values of
$\alpha=(0.5,10)$. Panel (b) is a magnification of panel (a) in the
vicinity of the origin.}  
\label{fig:tail_Elliptic_2}
\end{figure}

An interesting example is provided by the symmetric GH
distribution with parameters $\lambda=1.5$, $\delta=1$, $\mu=[0,0]$,
$\Delta_{12}=0.8$, and two different values of $\alpha$. The
difference between the resulting surfaces of $\Psi^{(l)}_{X|Y}$ when $\alpha
=0.5$ and $\alpha=10$, is displayed in panel (a) of Figure
\ref{fig:tail_Elliptic_2}. Panel (b) zooms in on the region close to
the origin. Although the two bivariate distributions have the same lower weak TDC
value of $\chi_{l}=0.90$, the former has greater conditional
probability in the lower-left quadrant than the latter. The values of
the proposed lower TDCs on the other
hand,  are instead $\Lambda^{(l)}_{X|Y}(1)=0.01$ and
$\Lambda^{(l)}_{X|Y}(1)=0.06$, respectively. (For $p=0.7$ these values
increase to $0.25$ and $0.18$, respectively.)

\section{Simulation Study}
\label{sim_Analysis}

The goal of this section is to assess the performance of maximum
likelihood estimation (MLE) in parametric modeling of the proposed TDCs
$\Lambda^{(l)}$ and  $\Lambda^{(u)}$ defined in \eqref{eq:Lambda_tail_l}.
The idea is to fit a parametric copula to the bivariate data sample,
and to then compute the underlying surface integrals. In a
real data scenario, this scheme would be applied to a bivariate
sample of residuals obtained after appropriate modeling of the
raw data. For example, in the financial returns case study of
Section \ref{CaseStudy}, it is applied to the pairs of innovations resulting
from (separate) ARMA-GARCH fits to each of two indices. This then
justifies approximating the likelihood function in the usual way as a
product of individual densities, as would be the case for an
independent sample.   

We simulate samples of size $n=500$ and $n=1000$ from a
particular bivariate copula, to which a MLE algorithm is then applied
to fit the parameters.  This operation is replicated 1,000 times, providing an equal number of
estimated values for each TDC. The following
summary performance measures are then empirically computed from these 1,000
replicates: the mean, standard deviation, and mean squared
error (MSE) defined as the squared bias plus variance. 

Our choice of couplas encompasses
three representative cases. The Gumbel copula, which has only one parameter
and where the lower TDC is different from the
upper; the $t$ copula, which has two parameters, but where the lower and
upper TDCs are equal; and the GH case with the parameters
specified as in the second row of Table \ref{tab:GH_Table}. The latter
copula is of particular interest because it represents  the asymmetric
situation where the TDCs are different if
computed based on $X|Y$ or $Y|X$, and also due to the fact that it has
a substantially larger number of parameters to fit. 

The results, presented in Table
\ref{tab:sim}, confirm the good performance of MLE; a consistent and
efficient estimation procedure. As expected, the GH case produces the highest MSE,
bias and standard deviation, especially at the lower sample size. The
stability and predictable asymptotic behavior of MLE-based estimates
then furnishes the possibility of using a parametric bootstrap scheme
to quantify the uncertainty in the estimates. Note that this would be
the most expedient way to obtain confidence intervals for the tail
dependence measures; analytical calculations involving the Hessian  of
the log-likelihood being infeasible due to the intractable nature of
the mapping from parameter space to $\Lambda$ space.

\section{Case Study: Tail Dependence of Financial Indices}
\label{CaseStudy}

\begin{figure}[h!]
	\centering	
	\subfloat[  ]{ \includegraphics[scale= 0.38]{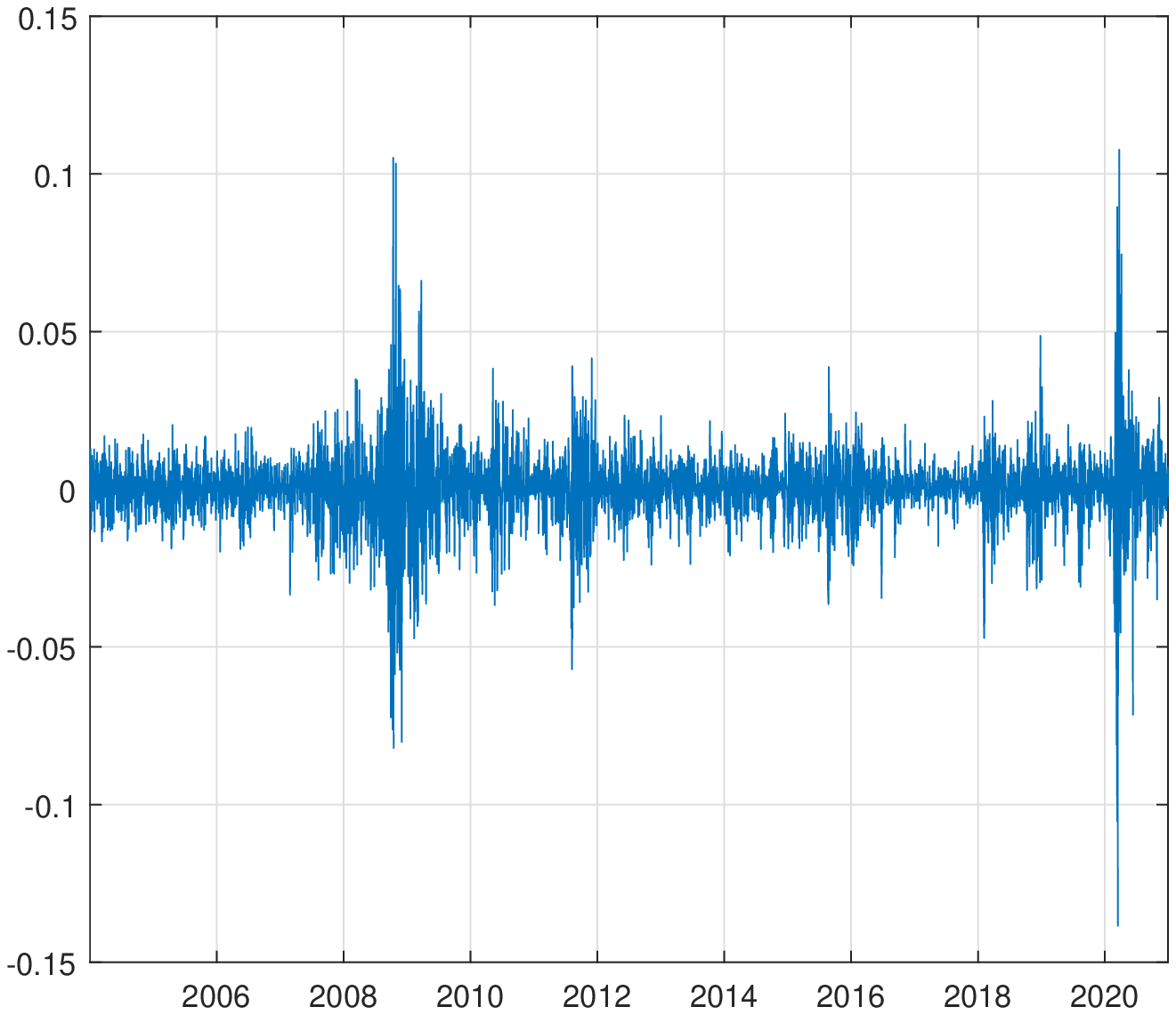}  } 
	\subfloat[  ]{ \includegraphics[scale= 0.38]{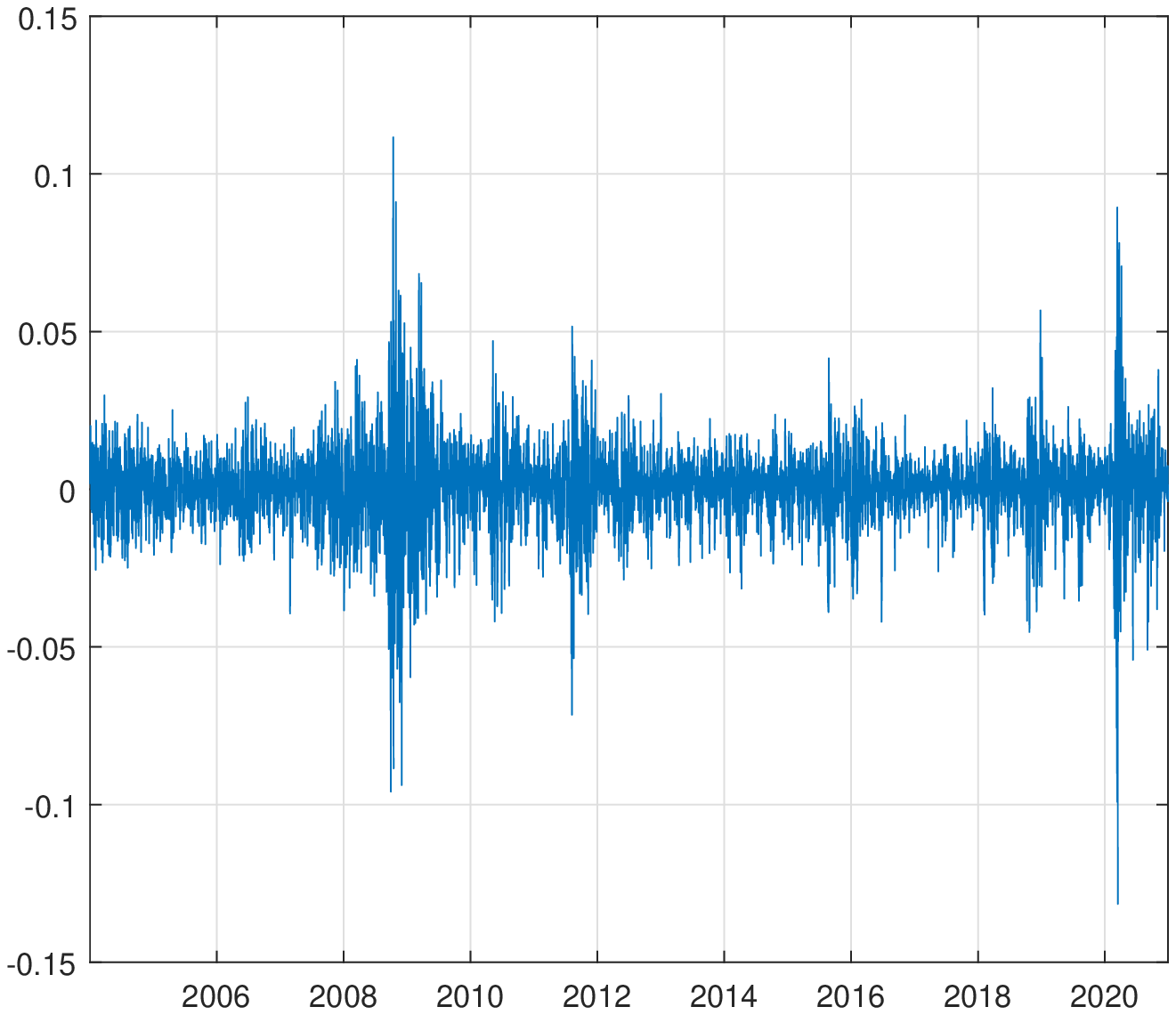}  }\\
	\subfloat[  ]{ \includegraphics[scale= 0.38]{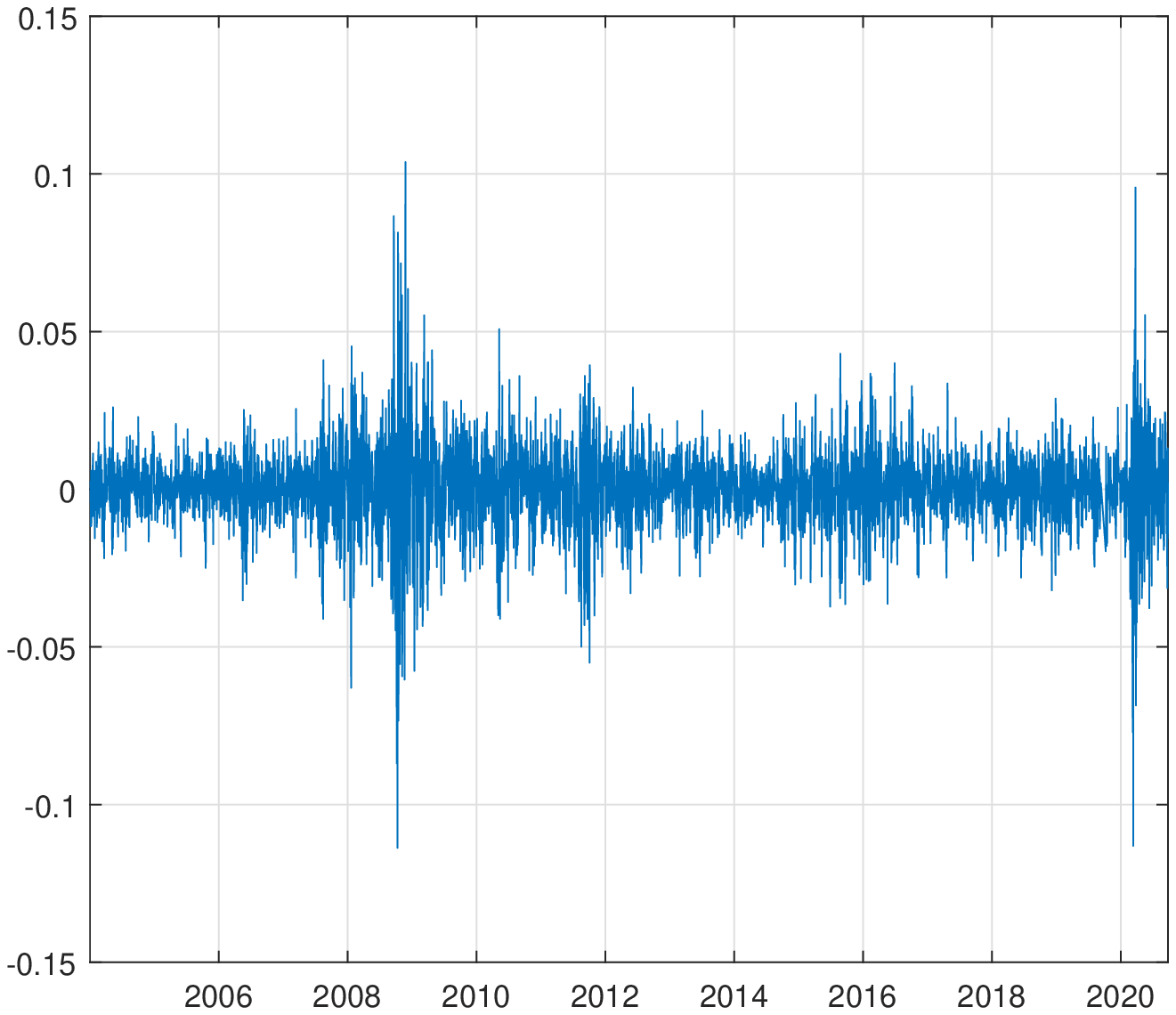} } 
	\subfloat[  ]{ \includegraphics[scale= 0.38]{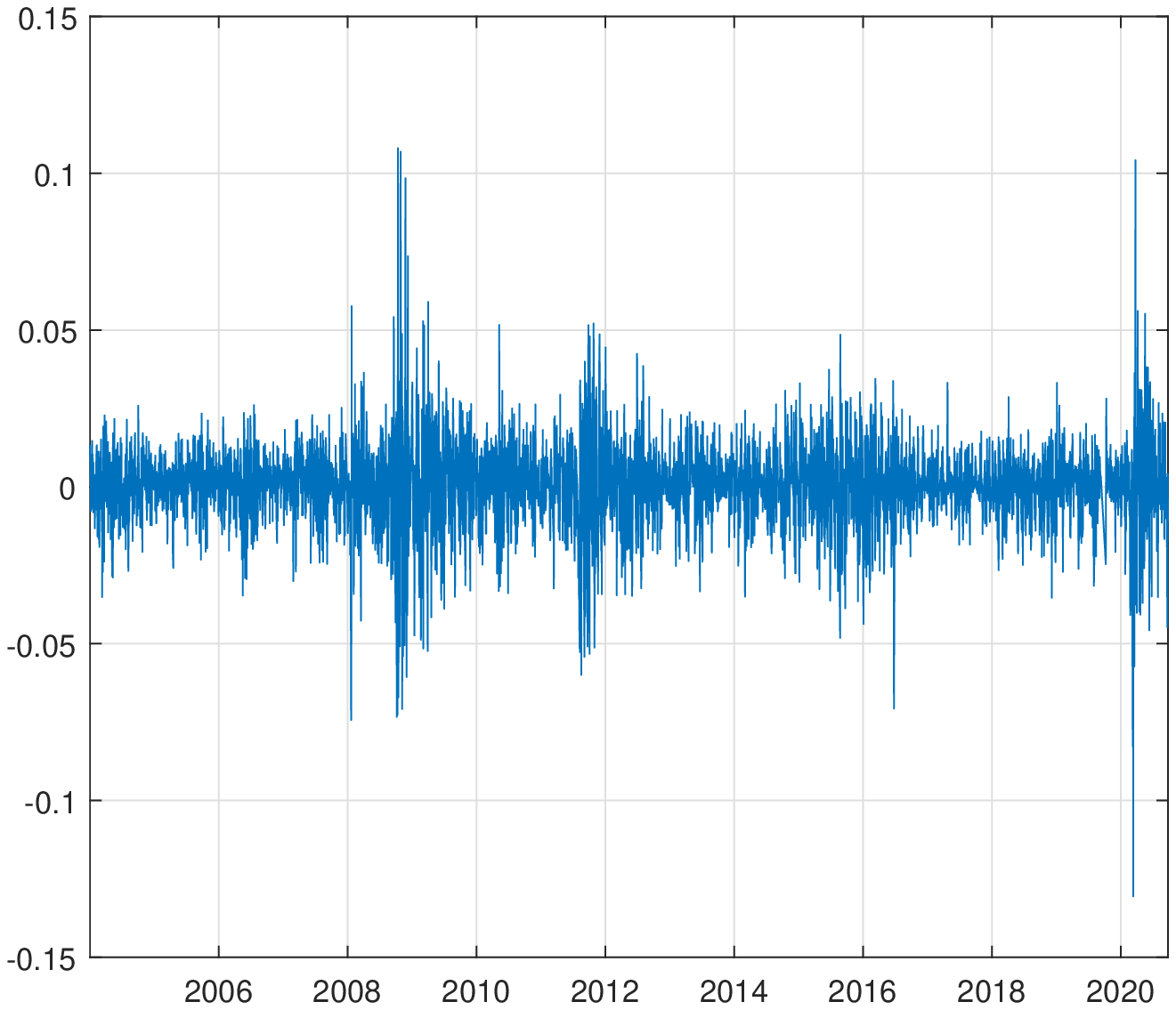} }\\
	\caption{Daily log returns for the indices: (a) Dow
            Jones Industrial Average, (b)Nasdaq Composite,
          (c) Fitse100, and (d) Dax30.}  
	\label{fig:Ret}
\end{figure}

In this section we illustrate the computation of the proposed TDCs
introduced in Section \ref{diffgeom_2} on a simple financial
application involving two bivariate datasets. The first consists of
daily log returns of the \textit{Dow Jones Industrial
  Average} (DJIA) and  \textit{Nasdaq Composite} indices, spanning the period
between from 04-Jan-2004  to 25-Sep-2020. The second dataset, spanning
the same period, is comprised of daily log returns of the two European indices \textit{Ftse100} and \textit{Dax30}, which represent, respectively,  the 100 companies by capital value listed on the London Stock Exchange and  the 30 companies by capital value listed on the Frankfurt Stock Exchange. The four series are plotted in Figure \ref{fig:Ret}. 

The goal is then to estimate $\Lambda^{(l)}$ and  $\Lambda^{(u)}$
between: (i) the two American indices, and (ii) the two European
indices. In order to clean rough market data from possible idiosyncratic
risk due to a particular index, we use a 500-point moving window fit.
That is, for each (trading) day in the time period ranging from 04-Jan-2006  to
25-Sep-2020, we fit separate univariate ARMA-GARCH models to the preceeding 500 data
points of each index (approximately two years). Appropriate lag orders for these  conditional mean
and variance  models, were selected based on the lowest
values of the information criterion BIC. All chosen models provided
plausible fits, as evidenced by the lack of serial correlation in both
the residuals and their squares. This then yields a bivariate vector,
denoted by $(X,Y)$,
of 500 innovations for each pair of indices at each day
in the time range.

At this point we apply the MCECM algorithm \cite{McNeil_2015} to fit
a bivariate GH distribution to $(X,Y)$, from which the four
TDCs $\Lambda^{(l)}_{X|Y},$ $\Lambda^{(l)}_{Y|X},$
$\Lambda^{(u)}_{X|Y},$ and $\Lambda^{(u)}_{Y|X}$ are computed (based
on the implied copula) via numerical integration on a structured
Cartesian mesh. To facilitate discussion, let $X$ and $Y$ represent
respectively the DJIA and Nasdaq (American case), and Ftse100 and
Dax30 (European case). The parametric bootstrap was used to
construct 95\% confidence bounds (using 5000 bootstrap replications). 

\begin{figure} 
	\centering	
	\includegraphics[ width=0.8\linewidth, height=0.3\textheight ]{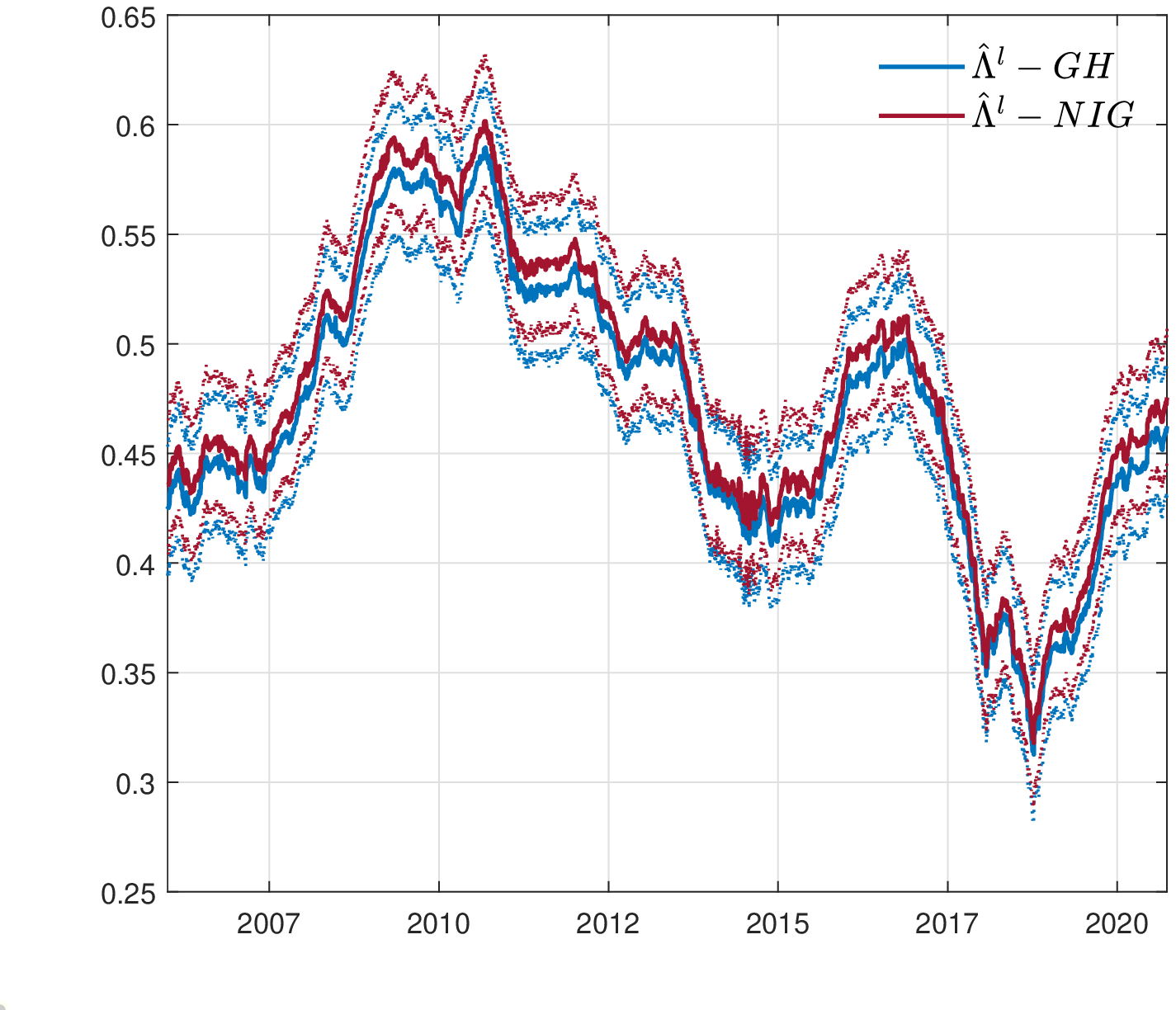}

	\caption{TDC $\Lambda^{(l)}_{X|Y}$ estimated from the
         ARMA-GARCH innovations fitted to each of the DJIA and Nasdaq returns, and based on either a
         GH copula (blue line) or NIG copula (red line). Corresponding
         95\% confidence bands are displayed as dotted lines.}
	
	\label{fig:Lambda_GH}

\end{figure}

\begin{figure} 
	\centering	
	\includegraphics[ width=0.8\linewidth, height=0.3\textheight ]{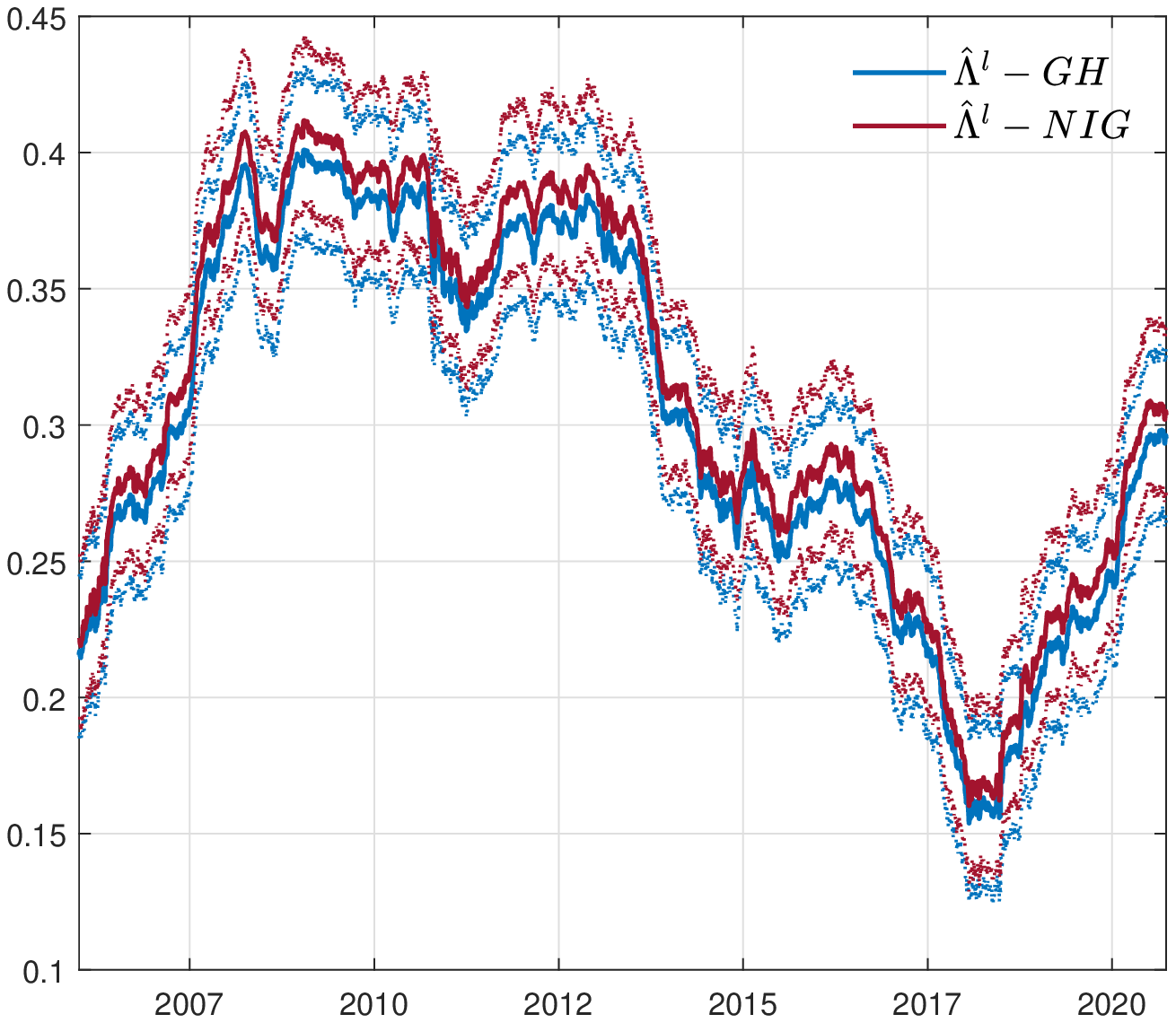}

	\caption{TDC $\Lambda^{(l)}_{X|Y}$ estimated from the
         ARMA-GARCH innovations fitted to each of the Ftse100 and Dax30 returns, and based on either a
         GH copula (blue line) or NIG copula (red line). Corresponding
         95\% confidence bands are displayed as dotted lines.}
	\label{fig:Lambda_GH_b}
	
\end{figure}

The estimated lower TDC  $\Lambda^{(l)}_{X|Y}$ for the two
American indices is displayed in  Figure \ref{fig:Lambda_GH}, and for the two
European indices in Figure \ref{fig:Lambda_GH_b}. The other three TDCs show
 very similar features and are not reported, but can be obtained from the authors upon request.
The estimates have a relatively narrow confidence band, as would be expected for 
parametric fits based on large samples,  and exhibit trajectories that
can be attributed to market conditions, with both lower and upper tail
dependence increasing during periods of higher volatility. As a
robustness check on the stability of our estimates against model
misspecification,  we replicated the procedure to estimate
$\Lambda^{(l)}$ and $\Lambda^{(u)}$ for the two American indices under
the hypothesis that innovations followed a  NIG distribution. As shown in Figure \ref{fig:Lambda_GH}, the
resulting estimates (red lines) are very close to those obtained under the
original GH assumption (blue lines). 

\begin{figure} 
	\centering		
    \subfloat[  ]{\includegraphics[scale= 0.4]{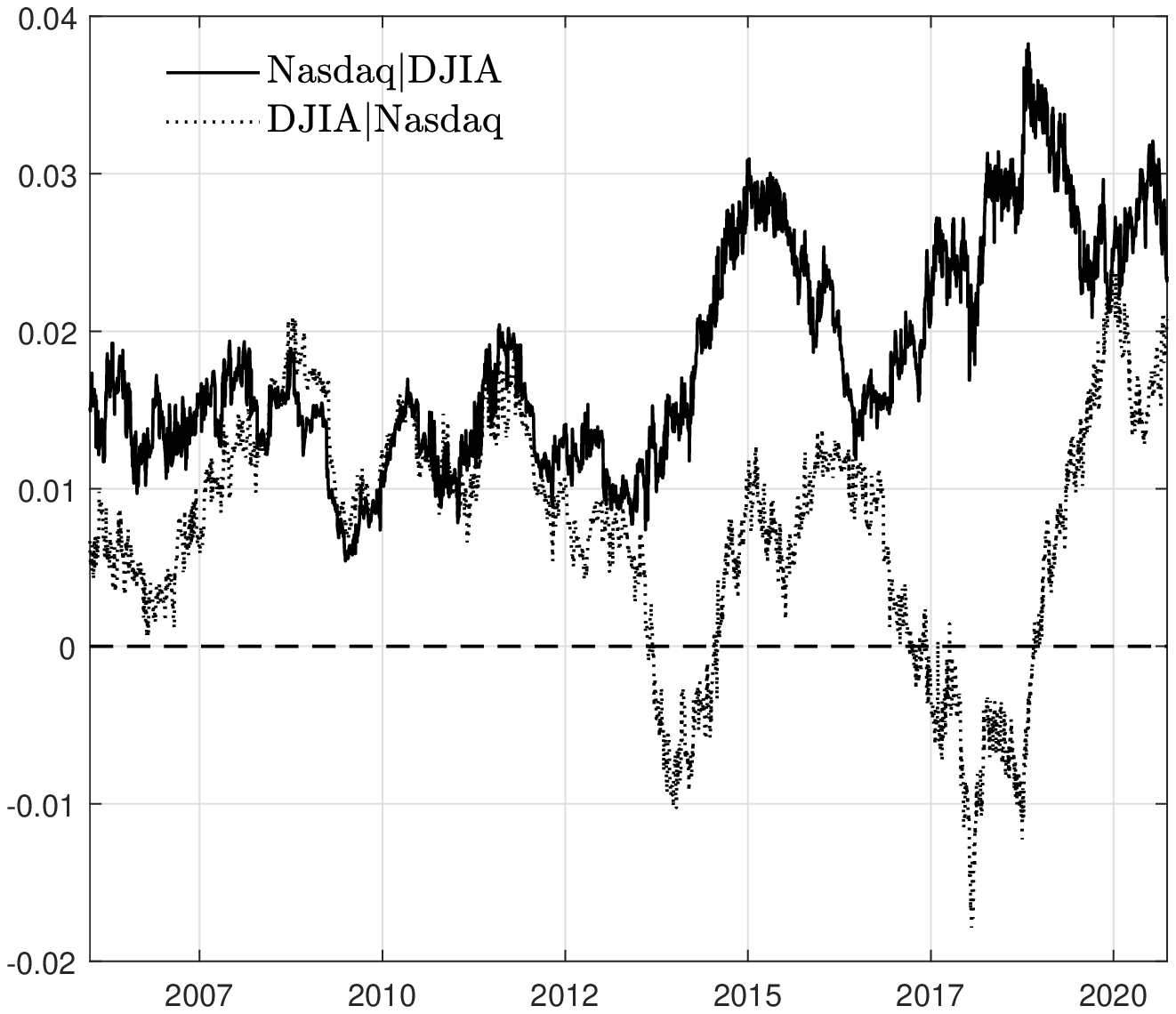}}
    \subfloat[  ]{\includegraphics[scale= 0.4]{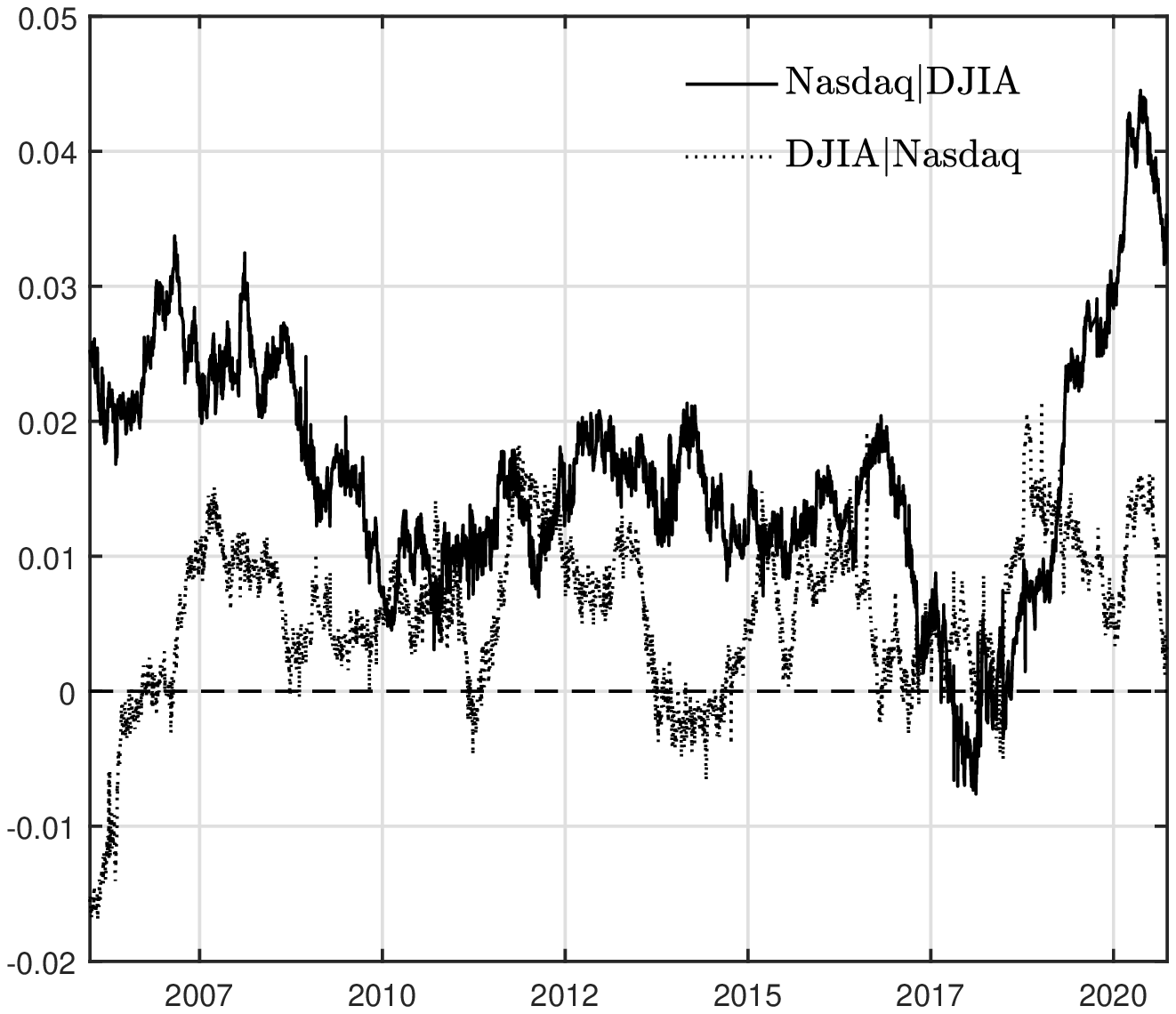}}\\
	\caption{ Panel (a):  The difference
          $\Lambda^{(l)}-\Lambda^{(u)}$ for the cases 
          Nasdaq$|$DJIA (solid) and DJIA$|$Nasdaq (dotted).
	Panel (b): The difference $\Lambda^{(l)}-\Lambda^{(u)}$ for
        the cases Dax30$|$Ftse100 (solid) and Ftse100$|$Dax30 (dotted).  } 
	\label{fig:Lambda_GH_2}	
\end{figure}

Differences  between the lower and upper TDCs, $\Lambda^{(l)}-\Lambda^{(u)}$, can be seen in Figure
\ref{fig:Lambda_GH_2}, for the American and European indices
in panels (a) and (b),
respectively. These differences range approximately from $-0.02$ to
$+0.045$, which can be interpreted as being not significant; however,
if we consider the  results in Table \ref{tab:Lambda_GH}, in
particular the second row related to the $GH_{2}$ distribution, we can
see that asymmetric structures can exist even with differences as
small as
$0.05$ units. Our results confirm, in general, the empirical finding
of \cite{Poon_2004} that lower tail dependence between stock returns
tends to be stronger than the upper.

\section*{Conclusions}
\label{Conclusion}

We have introduced alternative measures of global and tail
dependence between two random  variables. In the global case, the essence of the idea was
to measure the surface area of the conditional cumulative
probability in copula space, normalized
with respect to departures from the independence copula, and scaled by
the difference between the two boundary copulas of co-monotonicity
(positive dependence) and counter-monotonicity (negative
dependence). Two different normalization schemes lead to either a
measure of \emph{dependence} (bounded between $0$ and $1$) or of \emph{concordance} (bounded between $-1$ and $1$). The measures could be approached by cumulating
probability on either the lower
left or upper right domain of the copula space, leading to \emph{lower}
and \emph{upper} versions. 

The TDC has been similarly derived by altering
the surface integral of the global measure. In particular, we noted
that the product between the two partial derivatives of the
conditional cumulative probability surface is negative when the
conditional probability at a given point is higher than the corresponding value in the independence case. Furthermore, the absolute value of this
product increases as the copula converges toward the co-monotone case. We can then
bring out the tail dependence structure by simply weighting each
infinitesimal area element in the surface integral by means of  a
function of such a product of partial derivatives.

The measures have interesting features that make them competitive
in detecting dependence in regions of the support where
both random variables assume extreme values. In particular, they are able to differentiate asymmetric dependence with
respect to direction of conditioning,  resulting in a smoother and more refined
taxonomy  of global and tail dependence structures than that typically
delivered by the usual measures of dependence and concordance.

More investigation on estimation procedures could of course be carried
out; here we only considered the obvious parametric approach resulting
from maximization of the likelihood pertaining to a choice of copula model. A fully
non-parametric approach could in principle be achieved by estimating
an empirical copula, such as the Beta \cite{Segers_2017} or Checkerboard
\cite{Cuberos_2019}. Obvious computational difficulties with this route would likely
arise in the calculation of the necessary
surface integrals due to the necessity of performing numerical differentiation and integration. 

A set of possible extensions naturally emerge from this preliminary
work.  Generalizations of the measures to multi-dimensional random
vectors  would be of particular interest. Although the theory could in 
principle be easily adapted in this direction by considering
hyper-surface integrals, the (numerical) computational burden would increase
substantially. In this regard,  perhaps specialized numerical
integration algorithms that take into consideration non-uniformly
spaced grids could be brought to bear. It may also be fruitful to
study other representations of the measures, perhaps even through
characteristic functions or copula generating functions, which could
in turn lead to more efficient numerical algorithms.


\begin{appendix}

\section{Copula Theory - Basic results} \label{copulatheory}

In this section we summarize basic definitions and properties of
copulas; references are \cite{joe1997multivariate} and \cite{Nelsen2010}. 
Let $S_{1}$ and $S_{2}$ be two intervals in $\mathbb{R}$ and let $D$
be the planar region defined as their Cartesian product: $D = S_{1} \times S_{2}$. Let  $B=\left[ x_{1}, x_{2}\right] \times \left[ y_{1}, y_{2}\right] $ be a rectangular region in $D$ and $H: \mathbb{R}^{2} \supseteq D \rightarrow \mathbb{R}$ be a function of two variables. The $H$-volume of $B$ is  
$$V_{H}\left( B \right) = H\left( x_{2},y_{2} \right) + H\left(x_{1},y_{1}  \right) - H\left(x_{1},y_{2}  \right)-H\left( x_{2},y_{1} \right).$$
\begin{definition} \label{def:2increasing}
	The function $H$ is called \emph{2-increasing} if $V_{H}\left( B \right) \ge 0,$ $\forall B \in D$.
\end{definition}
\begin{definition} \label{def:grouded}
	Suppose that $\min \left(S_{1} \right)= a_{1}$ and $\min
        \left(S_{2} \right)= a_{2}$. Then $H$ is called grounded if 
\[
H\left( x,a_{2}\right)=0=H\left( a_{1},y\right), \text{ } \forall
\left( x , y \right) \in D.
\]
\end{definition}
\begin{lemma} \label{lemma:nondecreasing}
	Let $H$ be a 2-increasing function with domain $D$. If $H$ is also grounded then $H$ is also non-decreasing in each argument.  
\end{lemma}
\begin{definition} \label{def:margins}
	Suppose that $\max \left(S_{1} \right)= b_{1}$ and $\max \left(S_{2} \right)= b_{2}$.
	Then $H$ has margin functions $F:\mathbb{R} \supseteq S_{1}
        \rightarrow \mathbb{R}$ and $G:\mathbb{R} \supseteq S_{2}
        \rightarrow \mathbb{R}$ given by $F\left( x \right) = H\left( x , b_{2} \right)$ and
$G\left( x \right) = H\left( b_{1},y \right)$.
\end{definition}
\begin{definition}[Bivariate Copula]\label{def:2copula}	
	Let $I$ be the unit interval $\left[ 0 , 1\right]$ and
        $I^2=I\times I$ its Cartesian product. A bivariate copula $C$ is a function with domain $I^2$ such that:
	\begin{enumerate}
		\item $C$ is grounded and \textit{2-increasing}, and
		\item $ C\left( u,1\right)=u$, $\forall u \in I$ and  $C\left(1,v \right)=v$, $\forall  v \in I $. 
	\end{enumerate}
	
\end{definition}
Note the basic facts that: a copula is uniformly continuous in its domain; any convex linear combination of copulas is a copula. 
\begin{theorem} \label{th:C1deriv}
	Let $C(u,v)$ be a copula. Then the following hold.
	\begin{enumerate}
		\item For any $v$ in $I$ the partial derivative $\partial{C}/\partial{u}$ exists for almost all $u$ and $0 \le \partial{C}/\partial{u}  \le1$.
		
		\item For any $u$ in $I$ the partial derivative
                  $\partial{C}/\partial{v}$ exists for almost all $v$
                  and $0\le\partial{C}/\partial{v}\le 1$.
		\item The functions $u \rightarrow\partial{C}/\partial{v}$ exist and $v \rightarrow \partial{C}/\partial{u}$ are defined and non-decreasing almost everywhere on $I$.
	\end{enumerate}
\end{theorem}
%
\begin{theorem}[Sklar's Theorem] \label{th:Sklar}
	Let $H$ be a bivariate distribution function with margins $F$ and $G$. Then, there exists a copula $C$ such that for all $(x,y) \in \bar{ \mathbb{R} }^{2}$, $H\left( x , y \right) = C \left( F\left(x\right) , G\left(y\right) \right).$
	If $F$ and $G$ are continuous, then $C$ is unique. For every
        $\left(u,v\right)\in I^2$ we also have that $ C\left( u , v \right) = H \left( F^{-1}\left(u\right) , G^{-1}\left(v\right) \right).$
\end{theorem}
%
\begin{theorem}
	Let  $X$ and $Y$ be two continuous random variables with distribution functions $F$ and $G$, respectively, and joint distribution 
	function $H$. Let $C$ be the copula obtained as 
	$ C\left( u , v \right) = H \left( F^{-1}\left(u\right) , G^{-1}\left(v\right) \right).$
	Then $X$ and $Y$ are independent if and only if $C\left( u , v \right)=uv.$	
\end{theorem}
%
\begin{theorem}[Fr\'echet-Hoeffding Bounds] \label{th:FBbounds}
Define the functions $W: I^{2} \rightarrow I$ and $M: I^{2} \rightarrow I$ as,
$W\left( u,v \right) := \max \left( u+v-1,0\right)$, and 
$M\left( u,v \right) := \min \left( u , v \right)$.
Then, for every copula $C$ we have that
\[
  W\left( u,v \right) \le C\left( u,v \right) \le M\left( u,v \right),
  \qquad\text{for all }\left(u,v \right) \in I^{2}.
\]
\end{theorem}
%
\begin{theorem} \label{th:miTras}
	Let $X$ and $Y$ be continuous random variables with copula
        $C_{X,Y}$. If $\alpha$ and $\beta$ are strictly increasing
        functions on the supports of $X$ and $Y$, respectively, then $C_{\alpha(X),\beta(Y)} = C_{XY}$. Thus $C_{X,Y}$ is 
	invariant under strictly increasing transformations of $X$ and $Y$.
\end{theorem}
%
\begin{definition}[Measure of Dependence]
	\label{def:Dependence}
	A numeric measure $\delta$ of association between two continuous random variables $X$ and $Y$ whose copula is $C$, is a measure of \emph{dependence} if it satisfies the following properties (we write $\delta_{X,Y}$ or $\delta_{C}$ when needed): 
	\begin{enumerate}
		\item $\delta$ is defined for every pair $X$, $Y$ of continuous random variables;
		
		\item $0 \le \delta_{X,Y} \le 1$; 
		
		\item $\delta_{X,Y} = \delta_{Y,X}$
		
		\item $\delta_{X,Y} = 0$  if and only if $X$ and $Y$ are independent;
		
		\item $\delta_{X,Y} = 1$  if and only if each of $X$ and $Y$ is almost surely a strictly monotone function of the other;
		
		\item if $\alpha$ and $\beta$ are almost surely
                  strictly monotone functions on the supports of $X$
		and $Y$, respectively, then $\delta_{\alpha(X),\beta(Y)} = \delta_{X,Y}$
		
		\item if $\left\{( X_{n} ,Y_{n} )\right\}$ is a
                  bivariate sequence of continuous random variables with copulas $C_{n}$ , and if $\left\{ C_{n} \right\}$ converges pointwise to $C$, then $\lim_{n \to \infty } \delta_{ C_{n} } = \delta_{C} .$
	\end{enumerate}
\end{definition}
\noindent An example of a dependence measure is given by the  Schweizer and Wolff's $\sigma$:
\begin{align} \label{eq:Wolff}
   &\sigma = 12 \int_{I^{2}} |C(u,v) - uv| du dv.
\end{align}
\begin{definition}[Measure of Concordance]
	\label{def:Concordance}
	A numeric measure $\kappa$ of association between two continuous random variables $X$ and $Y$ whose copula is $C$, is a measure of \emph{concordance} if it satisfies the following properties (we write $\kappa_{X,Y}$ or $\kappa_{C}$ when needed): 
	\begin{enumerate}
		\item $\kappa$ is defined for every pair $(X,Y)$ of continuous random variables;
		
		\item $-1 \le \kappa_{X,Y} \le 1$; $\kappa_{X,X}=1$ and $\kappa_{X,-X}=-1$
		
		\item $\kappa_{X,Y} = \kappa_{Y,X}$
		
		\item if $X$ and $Y$ are independent, then $\kappa_{X,Y} = \kappa_{\Pi} = 0$;
		
		\item $\kappa_{-X,Y} = \kappa_{X,-Y}=- \kappa_{X,Y}$
		
		\item if $C_{1}$ and $C_{2}$ are copulas such that $C_{1} \prec C_{2}$ , then $\kappa_{C_{1}} \le \kappa_{C_{2}}$   
		
		\item if $\left\{( X_{n} ,Y_{n} )\right\}$ is a bivariate sequence of continuous random variables with copulas $C_{n}$ , and if $\left\{ C_{n} \right\}$ converges pointwise to $C$, then $\lim_{n \to \infty } \kappa_{ C_{n} } = \kappa_{C} .$
	\end{enumerate}
\end{definition}
Popular measures of concordance are Kendall's tau 	
\[
\tau_{C} = 4\int_{I^{2}} C(u,v)\text{ }dC(u,v)-1
	= 1-4\int_{I^{2}} \frac{\partial}{\partial u}C(u,v) \frac{\partial}{\partial v}C(u,v)\text{ } du dv,
\]
and Spearman's rho
	\[
	\rho_{C} = 12\int_{I^{2}} uv\text{ }dC(u,v)-3 
	= 12\int_{I^{2}} C(u,v)\text{ }dudv - 3
	= 12\int_{I^{2}} \left[ C(u,v)-uv \right] \text{ }dudv .
	\]	
Thus $\rho_{C}$ is a measure of "average distance" between the   distributions of $X$ and $Y$ (as represented by $C$)
	and independence (as represented  by the copula $\Pi$).

\section{Proofs} \label{AppendixB}

\subsection*{Theorem \ref{th:Asurf_disequality}}
%
\noindent The last equality in (\ref{eq:Asurf_disequality}) is  trivial: 
$$\int_{\mathbb{I}_{X|Y} } 1 \text{ }dS = \int_{0}^{1} \int_{0}^{1}  \sqrt{2} \text{ } du  dv=\sqrt{2}$$.
It is also easy to prove that $\int_{\mathbb{W}} 1 \text{ }dS = \int_{\mathbb{M}} 1 \text{ }dS$. 
Now, for the terms in the first integral
\begin{align}
    &  \mathbb{W}_{X|Y} &&= \frac{1}{ 2v }  \left( |u+v-1| +u+v-1 \right),\\
    &\frac{\partial  \mathbb{W}_{X|Y} }{\partial u} &&= \frac{1}{ 2v }  \cdot \left[ \text{sign}(u+v-1)+1 \right],\\
    &\frac{\partial  \mathbb{W}_{X|Y} }{\partial v} &&= \frac{1}{ 2v }  \cdot \left[ \text{sign}(u+v-1)+1  - \frac{|u+v-1| +u+v-1}{v}\right].
\end{align}
When $u+v+1<0$ then $\partial\mathbb{W}_{v} /\partial u=\partial\mathbb{W}_{v}/\partial v=0$, so we can write
\begin{align}
     &\int_{\mathbb{W}_{X|Y}} 1 \text{ }dS =
     \int_{0}^{1} \int_{0}^{1-u} 1 \  dv  du +
     \int_{0}^{1} \int_{1-u}^{1} \sqrt{\frac{v^{4} + v^{2} +(1-u)^{2} }{v^{4}}} \text{ }  dv du.
\end{align}
Consider now the second integral terms
\begin{align}
&  \mathbb{M}_{X|Y} &&=  \frac{1}{ 2v }    \left( u + v - |u-v| \right),\\
&\frac{\partial  \mathbb{M}_{X|Y} }{\partial u} &&= \frac{1}{ 2v}   \left[ -\text{sign}(u-v)+1 \right],\\
&\frac{\partial  \mathbb{M}_{X|Y} }{\partial v} &&= -\frac{1}{ 2v^{2} }  \left[ u+v -|u-v|\right] + \frac{1}{2v}\left[ 1+ \text{sign}(u-v) \right].
\end{align}
When $v<u$ then $\partial\mathbb{M}_{X|Y}/\partial u=\partial\mathbb{M}_{X|Y}/\partial v=0$, so we can write
\begin{align}
 &\int_{\mathbb{M}_{X|Y}} 1 \text{ }dS = 
 \int_{0}^{1} \int_{0}^{u} 1 \  dv  du +
 \int_{0}^{1} \int_{u}^{1} \sqrt{\frac{v^{4} + v^{2} +u^{2} }{v^{4}}}  dv du.
\end{align}
We just need to define the simple transformation $w=1-u$ in order to rewrite the surface integral for $\mathbb{W}_{X|Y}$ as
$$ \int_{0}^{1} \int_{0}^{1-w} 1 \  dv  dw + \int_{0}^{1} \int_{w}^{1}
\sqrt{\frac{v^{4} + v^{2} +w^{2} }{v^{4}}} \text{ }  dv dw,$$
which coincides with the integral for $\mathbb{M}_{X|Y}$. 
In order to prove that $\int_{ \mathbb{M}_{X|Y} }dS \ge \int_{ \Psi^{(l)}_{X|Y} }dS$, we observe that for each $(u_{0},v_{0})$, the distance  
$| \Psi^{(l)}_{Y|X}(u_{0},v_{0}) -\mathbb{I}_{X|Y}(u_{0},v_{0}) |$ 
is maximized with  $\Psi^{(l)}_{Y|X} = \mathbb{W}_{X|Y}$ or $\Psi^{(l)}_{Y|X} = \mathbb{M}_{X|Y}$. Given that the boundary $\partial\Omega_{1}$ is common to all the surfaces $\Psi^{(l)}_{Y|X}$, the inequality  must hold.
%

\subsection*{Theorem \ref{th:DepMeasure}}
    	\begin{enumerate}
    		\item If $X$ and $Y$ are continuous they admit a unique continuous copula $C$. The surface area of a  continuous function $f:\mathbb{R}: \rightarrow \mathbb{R}$ is well-defined.
    		
    		\item Relationships $0 \le \delta^{(l)}_{X|Y}\le 1$ follow directly from  inequalities in  Theorem \ref{th:ConditionalFHbound}.
    		
    		\item Property 3 is satisfied, in general, if $C$ is a symmetric copula.
    		
    		\item  $\delta^{(l)}_{X|Y}=0 $ if and only if $A(\Psi^{(l)}_{X|Y}) = A(\mathbb{I} )$, but since $\Phi_{X|Y}$ is a 
		minimal surface area, in order to have $A(\Psi^{(l)}_{X|Y}) = A(\mathbb{I})$ we must have $\Psi_{X|Y} = 
		\mathbb{I} $.
    		
    		\item  $\delta^{(l)}_{X|Y}=1 $ only if  $A(\Psi^{(l)}_{X|Y}) = A(\mathbb{W})$ or $A(\Psi^{(l)}_{X|Y}) = A(\mathbb{M} )$. 
		We know that  when $X$ and $Y$ are continuous "$Y$ is almost surely an increasing function of $X$" if 
		and only if the copula of $X$ and $Y$ is $\mathbb{M}$, and "$Y$ is almost surely a decreasing function of $X$" if
		and only if the copula of $X$ and $Y$ is $\mathbb{W}$. Thus, $\delta^{(l)}_{X|Y}=1 $ if and only if each  $X$ and
		$Y$ is almost surely a strictly monotone function of the other.
    		
    		\item Follows directly form theorem \ref{th:miTras}.
    		
    		\item This is equivalent to proving that 
    		$$ \lim_{n \to \infty } \int_{0}^{1} \int_{0}^{1} \sqrt{1+ ||\nabla{\Psi^{(l)}_{n,v}}||^{2}} \  dv  du 
		= \int_{0}^{1} \int_{0}^{1} \sqrt{1+ ||\nabla{\Psi^{(l)}_{v}}||^{2}} \  dv  du,$$
		where $\Psi^{(l)}_{X|Y,n}:=C_{n}/v$, which follows from the fact that a continuous copula is uniformly 	continuous. 
    	\end{enumerate}

\section{Generalized Hyperbolic Distribution and Other Copulas} \label{AppendixC}
The bivariate Generalized Hyperbolic (GH) distribution, due to
\cite{BN77}, is defined as follows. 	Let $\Delta$ be a real-valued,
positive semidefinite  $\left(2 \times 2 \right)$ matrix and let   $
\lambda \in \mathbb{R}$, $\alpha, \delta \in \mathbb{R}_{+}$, and
$\beta, \mu \in \mathbb{R}^{2}$ be a
    	set of parameters satisfying  the following alternative constraints:
         \begin{align*}
        	   & \delta  \ge 0,0 \le \sqrt{\beta'\Delta \beta} < \alpha,\hspace{0.4cm}\text{ if } \lambda >0,\\
        	   &\delta  > 0, 0\le\sqrt{\beta'\Delta \beta} < \alpha, \hspace{0.4cm}\text{ if }  \lambda =0,\\
        	   &\delta  >0, 0\le\sqrt{\beta'\Delta \beta} \le \alpha, \hspace{0.4cm} \text{ if } \lambda <0.
         \end{align*}
        The GH density of the bivariate continuous random vector  $Z=[X,Y]$ is then defined as:
        \begin{align} \label{dMGH}
    	   h\left(z; \lambda, \alpha,\beta, \delta,\mu,\Delta \right)& = \frac{\left(\alpha^{2}-\beta'\Delta\beta \right)^{\frac{\lambda}{2}}}
    	      {\left(2\pi\right)^{\frac{2}{2}}\sqrt{|\Delta|}\text{ } \alpha^{\lambda-\frac{2}{2}} \delta^{\lambda} K_{\lambda}\left( \delta 
	      \sqrt{\alpha^{2} - \beta'\Delta\beta}\right)} \nonumber \\   
	   & \times \left( \left(z-\mu\right)'\Delta^{-1}\left(z-\mu\right)+\delta^{2}    \right)^{\frac{\left(\lambda-\frac{2}{2}\right)}{2}}  	        \nonumber \\
	   & \times K_{\lambda-1}\left( \alpha \sqrt{ \left(z-\mu\right)'\Delta^{-1}\left(z-\mu\right)+\delta^{2} } \right) e^{\beta' \left( z-  \mu\right)},
        \end{align}
         where  $|\Delta|$ is the determinant of $\Delta$, and $K_{\nu}$ is the modified Bessel function of the third kind with index $\nu$.
         The GH class contains important distribution families: the \textit{multivariate Normal Inverse Gaussian} (NIG) for $\lambda=-1/2$, the  \textit{multivariate Variance-Gamma} (VG)  if 
         $\lambda>0$  and  $\delta \to \infty$; in the case in which $\beta=\left[0,0 \right]$, the random vector $Z$ is elliptically contoured and its distribution is called multivariate 
         symmetric GH. The  \textit{multivariate  scaled and shifted t-distribution} with $-2\lambda$ degrees of freedom is part of  the symmetric sub-class, and is attained  for $\lambda<0$ and $
         \alpha \rightarrow 0$. The \textit{Multivariate  Normal}
         distribution is the limiting case of $\alpha \rightarrow
         \infty$, $\delta \rightarrow \infty$, and $\delta/\alpha 
         \rightarrow \sigma^{2} < \infty$.
A particular GH distribution will induce a specific (implied) copula, which in
general can only be numerically computed. \\

\noindent Functional forms for other
copula used in this paper are as follows.

\begin{itemize}	
	\item Fr\'echet Copula,  \cite{Frechet_1958}
	\begin{equation}
	C(u,v;\alpha,\beta) = \alpha M(u,v) +(1-\alpha-\beta) \Pi(u,v) +\beta W(u,v)
	\end{equation}
	for $\alpha \beta \in [0,1]$ and $\alpha + \beta \le 1$
		
	\item Mardia Copula, \cite{Mardia_1967}
	\begin{equation}
	C(u,v;\theta) = \frac{ \theta^{2} \left( 1 + \theta \right) }{2} M(u,v) + \left( 1 - \theta^{2} \right)  \Pi(u,v) + \frac{ \theta^{2} 
		\left( 1 + \theta \right) }{2}W(u,v)
	\end{equation}
	for $\theta \in [-1,1]$
	
	\item Cuadras-Aug\'e  Copula,  \cite{Cuadras1981}
	\begin{equation}
	C(u,v;\theta) =	 \left[ \min(u,v)^{\theta} \right]^{\theta} \left[ u v \right]^{1 - \theta}
	\end{equation}
	for $\theta \in [0,1]$
	
	\item Gumbel-Hougaard Copula,  \cite{Gumbel_1960b} \cite{Hougaard_1986}
	\begin{align}
	& C(u,v;\theta) =	\exp\left[  - \left[  \left(- \ln{u}^{\theta} \right) +  \left(- \ln{v}^{\theta} \right)  \right]^{ \frac{1}{\theta} } \right]
	\end{align}
	for $\theta \in [-1,1]$
		
	\item Ali-Mikhail - Haq  Copula, \cite{HutchinsonLai_1990}
	\begin{align}
	& C(u,v;\theta) =	\frac{uv}{1- \theta \left(  1-u \right)\left(  1-v \right) }
	\end{align}
	for $\theta \in [-1,1]$
	
	\item Clayton Copula,  \cite{Clayton_1978}
	\begin{align}
	& C(u,v;\theta) = \max{\left( u^{-\theta} + v^{-\theta} -1,0 \right)^{-\frac{1}{\theta}} }
	\end{align}
	for $\theta \in [-1,0) \cup (0,\inf) $
	
	\item Frank Copula, \cite{Frank_1979}
	\begin{align}
	& C(u,v;\theta) =	- \frac{1}{\theta} \ln{\left[  1 + \frac{ \left( e^{-\theta u}-1 \right) \left( e^{-\theta v}-1 \right) }{ e^{-\theta} -1 } 
	\right] }
	\end{align}
	for $\theta \in \mathbb{R}$

	\item Marshall and Olkin Copula,  \cite{MarshallOlkin_1967a} and \cite{MarshallOlkin_1967b}
	\begin{align}
	& C(u,v;\theta) =	\min{ \left( u^{1-\alpha} v, uv^{1-\beta}\right) }
	\end{align}
	for $\alpha >0$ and $\beta<1$

\end{itemize}

\end{appendix}

\section*{Tables}
\label{Tables}

\begin{table}[h!] 
\centering
	 \scalebox{1}{
 
	\begin{tabular}{ ccccccccccc } \hline \\ 
		Copula & $\lambda$ & $\alpha$ & $\beta_{1}$& $\beta_{1}$ & $\delta$ & $\mu_{1}$ & $\mu_{1}$ & $\Delta_{1,1}$ & $\Delta_{1,2}$ & $\Delta_{2,2}$   
		\\ \hline \\ 
		GH$_{1}$  & 1.50 & 1.10 & 0.00 & 0.00 & 1 & 0 & 0 & 2.29 & 2.06 & 2.29\\  
		GH$_{2}$  & 1.50 & 0.80 & -0.40 & 0.30 & 1 & 0 & 0 & 2.29 & 2.06 & 2.29\\  
		GH$_{3}$  & 1.00 & 1.30 & 0.00 & 0.00 & 1 & 0 & 0 & 1.77 & 1.57 & 1.96 \\  
		GH$_{4}$  & 1.00 & 1.30 & -0.40 & -0.30 & 1 & 0 & 0 & 3.43 & 0.69 & 0.43  \\  
		NIG$_{1}$  & -0.50 & 1.20 & 0.00 & 0.00 & 1 & 0 & 0 & 1.77 & 1.57 & 1.96  \\  
		NIG$_{2}$  & -0.50 & 1.20 & -0.40 & -0.30 & 1 & 0 & 0 & 1.77 & 1.57 & 1.96  \\  
		NIG$_{3}$  & -0.50 & 1.20 & -0.40 & -0.30 & 1 & 0 & 0 & 3.43 & 0.69 & 0.43  \\  
		NIG$_{4}$  & -0.50 & 1.20 & -0.40 & 0.30 & 1 & 0 & 0 & 3.43 & 0.69 & 0.43  \\ 
		VG$_{1}$  & 0.80 & 1.30 & 0.00 & 0.00 & 0 & 0 & 0 & 1.77 & 1.57 & 1.96 \\  
		VG$_{2}$  & 0.80 & 1.30 & -0.40 & -0.30 & 0 & 0 & 0 & 1.77 & 1.57 & 1.96  \\  
		VG$_{3}$  & 0.50 & 1.10 & -0.40 & -0.30 & 0 & 0 & 0 & 1.77 & 1.57 & 1.96  \\  
		VG$_{4}$  & 0.50 & 1.10 & -0.40 & -0.30 & 0 & 0 & 0 & 3.43 & 0.69 & 0.43  \\  
	\end{tabular} 
	}
		\caption{Selected bivariate GH distributions.}   
	\label{tab:GH_Table} 
\end{table}
	
\begin{table}[h!] 
\centering
	 \scalebox{1}{

	\begin{tabular}{ccccccccc} \hline \\ 
		Copula  & $\kappa^{(l)}_{X|Y}$ & $\kappa^{(l)}_{Y|X}$ & $\kappa^{(u)}_{X|Y}$ & $\kappa^{(u)}_{Y|X}$ & $\tau$ & $\rho$ \\ \hline \\ 
		GH$_{1}$  & 0.85 & 0.85 & 0.85 & 0.85 & 0.71 & 0.87  \\  
		GH$_{2}$  & 0.85 & 0.86 & 0.83 & 0.82 & 0.70 & 0.86  \\
		GH$_{3}$  & 0.78 & 0.78 & 0.78 & 0.78 & 0.64 & 0.81  \\  
		GH$_{4}$  & 0.71 & 0.71 & 0.59 & 0.58 & 0.53 & 0.70  \\  
		NIG$_{1}$  & 0.78 & 0.78 & 0.78 & 0.78 & 0.64 & 0.81  \\  
		NIG$_{2}$  & 0.89 & 0.89 & 0.80 & 0.80 & 0.71 & 0.87  \\  
		NIG$_{3}$  & 0.67 & 0.67 & 0.53 & 0.52 & 0.47 & 0.63  \\  
		NIG$_{4}$  & 0.55 & 0.54 & 0.45 & 0.47 & 0.39 & 0.53  \\  
		VG$_{1}$  & 0.78 & 0.78 & 0.78 & 0.78 & 0.64 & 0.80  \\  
		VG$_{2}$  & 0.91 & 0.91 & 0.82 & 0.82 & 0.76 & 0.90  \\  
		VG$_{3}$  & 0.94 & 0.94 & 0.84 & 0.84 & 0.80 & 0.92  \\  
		VG$_{4}$  & 0.77 & 0.79 & 0.59 & 0.55 & 0.58 & 0.74  \\     
	\end{tabular} 
	}
		\caption{Values of upper and lower concordance measure $\kappa$,
          Kendall's $\tau$, and Spearman's $\rho$,  computed for the
          implied GH copulas of Table~\ref{tab:GH_Table}.} 
	\label{tab:kappa_GH} 
\end{table} 
   
   \newpage 
\begin{table}[h!]
\centering
     \scalebox{1}{
    \begin{tabular}{ccccccccccc}
    \hline
    Copula & \multicolumn{2}{c}{ $\Lambda^{(l)}_{X|Y}$ } & \multicolumn{2}{c}{ $\Lambda^{(l)}_{Y|X}$ } & \multicolumn{2}{c}{ $\Lambda^{(u)}_{X|Y}$  } & 
    \multicolumn{2}{c}{    $\Lambda^{(u)}_{Y|X}$  } & $\lambda_{l}$& $\lambda_{u}$\\ \hline \\
                          p                & 1       & 0.7     & 1          & 0.7       & 1              & 0.7           & 1               & 0.7  &0&0         \\ \hline \\
                       GH$_{1}$   & 0.19  & 0.36   &  0.19    & 0.36     & 0.19         & 0.36         & 0.19          & 0.36   &0&0           \\
                       GH$_{2}$   & 0.23  & 0.40   &  0.19    & 0.35     & 0.13         & 0.29         & 0.19          & 0.35   &1&1               \\ 
                       GH$_{3}$   & 0.13  & 0.28   &  0.13    & 0.28     & 0.13         & 0.28         & 0.13          & 0.28   &0&0               \\
                       GH$_{4}$   & 0.12  & 0.27   &  0.10    & 0.25     & 0.01         & 0.06         & 0.02          & 0.10   &0&0               \\
                       NIG$_{1}$  & 0.14  & 0.30   &  0.14    & 0.30     & 0.14         & 0.30         & 0.14          & 0.30   &0&0              \\
                       NIG$_{2}$  & 0.37  & 0.52   &  0.36    & 0.51     & 0.09         & 0.23         & 0.09          & 0.24   &0&0               \\
                       NIG$_{3}$  & 0.16  & 0.30   &  0.13    & 0.27     & 0.01         & 0.06         & 0.02          & 0.10   &0&0               \\
    		  NIG$_{4}$  & 0.09  & 0.23   &  0.05    & 0.15     & 0.01         & 0.05         & 0.03          & 0.13    &0&0          \\  
                       VG$_{1}$   & 0.14  & 0.30   &  0.14    & 0.30      & 0.14        &  0.30        & 0.18          & 0.32    &0&0            \\
                       VG$_{2}$  & 0.33   & 0.50   &  0.33    & 0.50      & 0.09        & 0.24         & 0.09          & 0.25    &0&0             \\ 
                       VG$_{3}$  & 0.42  &  0.58   &  0.38    & 0.56      & 0.09        &   0.24       & 0.11           &  0.26  &0&0             \\
    		  VG$_{4}$  & 0.11  & 0.28     &  0.15    & 0.32     & 0.01         &   0.06       & 0.01          &  0.09  &0&0             \\
    \end{tabular}
    }
         \caption{Values of upper and lower tail dependence measure $\Lambda$ computed for
       the implied GH copulas of Table~\ref{tab:GH_Table}, and for two choices of the focus 
     parameter $p$. The last two columns list also the upper and lower
     strong TDC $\lambda$. Results are obtained with numerical integration on a uniform mesh with $10^{6}$ points. } 
    \label{tab:Lambda_GH} 
\end{table}

\newpage
\begin{table}[hbt!]
\centering
         \scalebox{1}{
	
	\begin{tabular}{ ccccccccc } \hline \\ 
		Copula & Param & $\kappa^{(l)}_{X|Y}$ & $\kappa^{(l)}_{Y|X}$ & $\kappa^{(u)}_{X|Y}$ & $\kappa^{(u)}_{Y|X}$ & $\tau$ & $\rho$ \\ \hline \\ 
		Frechet &  0.3,  0.7 & -0.40 & -0.40 & -0.40 & -0.40 & -0.39 & -0.41  \\  
		Frechet &  0.5,  0.5 & 0.00 & 0.00 & 0.00 & 0.00 & 0.00 & 0.01  \\  
		Frechet &  0.7,  0.3 & 0.40 & 0.40 & 0.40 & 0.40 & 0.40 & 0.39  \\    
		Gumbel &  4 & 0.84 & 0.84 & 0.91 & 0.91 & 0.75 & 0.90  \\  
		Gumbel &  6 & 0.91 & 0.91 & 0.96 & 0.96 & 0.83 & 0.95  \\  
		Gumbel &  10 & 0.96 & 0.96 & 0.99 & 0.99 & 0.90 & 0.97  \\  
		Clayton &  1 & 0.55 & 0.55 & 0.36 & 0.36 & 0.34 & 0.47  \\  
		Clayton &  2 & 0.75 & 0.75 & 0.53 & 0.53 & 0.50 & 0.67  \\  
		Clayton &  5 & 0.92 & 0.92 & 0.73 & 0.73 & 0.72 & 0.87  \\  
		Clayton &  10 & 0.97 & 0.97 & 0.84 & 0.84 & 0.83 & 0.95  \\  
		Clayton &  30 & 1.00 & 1.00 & 0.94 & 0.94 & 0.94 & 0.98  \\    
		Frank &  1.5 & 0.20 & 0.20 & 0.20 & 0.20 & 0.17 & 0.23  \\  
		Frank &  5 & 0.55 & 0.55 & 0.55 & 0.55 & 0.46 & 0.63  \\  
		Frank &  15 & 0.84 & 0.84 & 0.84 & 0.84 & 0.76 & 0.92  \\  
		Ali &  0.5 & 0.17 & 0.17 & 0.15 & 0.15 & 0.13 & 0.18  \\  
		Ali &  1 & 0.55 & 0.55 & 0.36 & 0.36 & 0.34 & 0.47  \\  
		Mardia &  -0.9 & -0.71 & -0.71 & -0.71 & -0.71 & -0.68 & -0.74  \\  
		Mardia &  0.9& 0.71 & 0.71 & 0.71 & 0.71 & 0.68 & 0.72  \\  
		Cuadras &  0.5 & 0.35 & 0.35 & 0.41 & 0.41 & 0.34 & 0.42  \\  
		Cuadras &  0.8 & 0.67 & 0.67 & 0.74 & 0.74 & 0.67 & 0.74  \\
		Gaussian &  0.5 & 0.44 & 0.44 & 0.44 & 0.44 & 0.33 & 0.47  \\  
		Gaussian &  0.7 & 0.62 & 0.62 & 0.62 & 0.62 & 0.49 & 0.67  \\  
		Gaussian &  0.9 & 0.85 & 0.85 & 0.85 & 0.85 & 0.71 & 0.88  \\  
		Gaussian &  0.95 & 0.92 & 0.92 & 0.92 & 0.92 & 0.80 & 0.93  \\    
		Marshall &  0.5,  0.5 & 0.35 & 0.35 & 0.41 & 0.41 & 0.34 & 0.42  \\  
		Marshall &  0.6,  0.1 & 0.13 & 0.09 & 0.12 & 0.15 & 0.10 & 0.12  \\  
		Marshall &  0.7,  0.9 & 0.68 & 0.62 & 0.71 & 0.76 & 0.65 & 0.72  \\  
		$t$ &  0.50,  4 & 0.42 & 0.42 & 0.42 & 0.42 & 0.33 & 0.46  \\  
		$t$ &  0.90,  4 & 0.85 & 0.85 & 0.85 & 0.85 & 0.71 & 0.87  \\  
		$t$ &  0.95, 4 & 0.92 & 0.92 & 0.92 & 0.92 & 0.80 & 0.93  \\
		$t$ &  0.50,  2 & 0.42 & 0.42 & 0.42 & 0.42 & 0.33 & 0.44  \\  
		$t$ &  0.90,  2 & 0.85 & 0.85 & 0.85 & 0.85 & 0.71 & 0.86  \\  
		$t$ &  0.95, 2 & 0.92 & 0.92 & 0.92 & 0.92 & 0.80 & 0.92  \\  
		$t$ &  0.50,  1 & 0.41 & 0.41 & 0.41 & 0.41 & 0.34 & 0.42  \\  
		$t$ &  0.90,  1 & 0.83 & 0.83 & 0.83 & 0.83 & 0.71 & 0.83  \\  
		$t$ &  0.95, 1 & 0.90 & 0.90 & 0.90 & 0.90 & 0.80 & 0.90  \\
		
	\end{tabular} 
	}
	
	 	\caption{Values of upper and lower  concordance measure $\kappa$,
          Kendall's $\tau$, and Spearman's $\rho$,
          computed   for the
          selection of non-GH copulas discussed in Appendix~\ref{AppendixC}. Results are obtained with numerical integration on a structured Cartesian 
	mesh with $10^{6}$ points.   } 
	\label{tab:kappa} 
\end{table} 
	\newpage
\begin{table}[hbt!] 
\centering
      \scalebox{0.85}{
	\begin{tabular}{ ccccccccc } \hline \\ 
		Copula & Param & $\Lambda^{(l)}_{X|Y}(1.0)$ & $\Lambda^{(l)}_{Y|X}(1.0)$ & $\Lambda^{(u)}_{X|Y}(1.0)$ & $\Lambda^{(u)}_{Y|X}(1.0)$ & $
		\lambda_{l}$ & $\lambda_{u}$ \\ \hline \\ 
		Frechet  &  0.3,  0.7 & 0.03 & 0.03 & 0.03 & 0.03 & 0.3 & 0.3  \\  
		Frechet  &  0.5,  0.5 & 0.13 & 0.13 & 0.13 & 0.13 & 0.5 & 0.5  \\  
		Frechet  &  0.7,  0.3 & 0.34 & 0.34 & 0.34 & 0.34 & 0.7 & 0.7  \\  
		Gumbel   &  1        & 0.00 & 0    & 0.09 & 0.09 & 0 & 0.41  \\  
		Gumbel   &  4        & 0.07 & 0.07 & 0.48 & 0.48 & 0 & 0.81  \\  
		Gumbel   &  6        & 0.15 & 0.15 & 0.62 & 0.62 & 0 & 0.88  \\  
		Gumbel   &  10       & 0.29 & 0.29 & 0.77 & 0.77 & 0 & 0.93  \\  
		Clayton  &  1        & 0.14 & 0.14 & 0    & 0    & 0.50 & 0  \\  
		Clayton  &  2        & 0.32 & 0.32 & 0    & 0    & 0.71 & 0  \\  
		Clayton  &  5        & 0.61 & 0.61 & 0    & 0    & 0.87 & 0  \\  
		Clayton  &  10       & 0.78 & 0.78 & 0.01 & 0.01 & 0.93 & 0  \\  
		Clayton  &  30       & 0.92 & 0.92 & 0.02 & 0.02 & 0.98 & 0  \\  
		Frank    &  5        & 0    & 0    & 0    & 0    & 0    & 0  \\  
		Frank    &  15       & 0.01 & 0.01 & 0.01 & 0.01 & 0    & 0  \\    
		Cuadras  &  0.2      & 0    & 0    & 0.01 & 0.01 & 0    & 0.20  \\  
		Cuadras  &  0.5      & 0    & 0    & 0.13 & 0.13 & 0    & 0.50  \\  
		Cuadras  &  0.8      & 0.04 & 0.04 & 0.51 & 0.51 & 0    & 0.80  \\      
		Marshall &  0.1,  0.6 & 0.00 & 0.00 & 0.17 & 0.00 & 0    & 0.10  \\  
		Marshall &  0.5,  0.5 & 0.00 & 0.00 & 0.13 & 0.13 & 0    & 0.50  \\  
		Marshall &  0.6,  0.1 & 0.00 & 0.00 & 0.00 & 0.17 & 0    & 0.10  \\  
		Marshall &  0.7,  0.9 & 0.01 & 0.15 & 0.65 & 0.30 & 0    & 0.70  \\  
		Gaussian &  0.700 & 0.04 & 0.04 & 0.04 & 0.04 & 0.00 & 0.00  \\  
		Gaussian &  0.900 & 0.15 & 0.15 & 0.15 & 0.15 & 0.00 & 0.00  \\  
		Gaussian &  0.950 & 0.25 & 0.25 & 0.25 & 0.25 & 0.00 & 0.00  \\  
		Gaussian &  0.990 & 0.53 & 0.53 & 0.53 & 0.53 & 0.00 & 0.00  \\
		$t$ &  0.500,  2 & 0.07 & 0.07 & 0.07 & 0.07 & 0.42 & 0.42  \\  
		$t$ &  0.900,  2 & 0.33 & 0.33 & 0.33 & 0.33 & 0.73 & 0.73  \\  
		$t$ &  0.950,  2 & 0.46 & 0.46 & 0.46 & 0.46 & 0.81 & 0.81  \\ 
		$t$ &  0.500,  1 & 0.11 & 0.11 & 0.11 & 0.11 & 0.56 & 0.56  \\  
		$t$ &  0.900,  1 & 0.42 & 0.42 & 0.42 & 0.42 & 0.80 & 0.80  \\  
		$t$ &  0.950,  1 & 0.54 & 0.54 & 0.54 & 0.54 & 0.86 & 0.86  \\ 
	\end{tabular} 
	}
        \caption{Values of upper and lower tail dependence measure $\Lambda$ computed for
       the non-GH copulas discussed in Appendix~\ref{AppendixC}. The
       value of the focus parameter is set at $p=1$. The last two columns list also the upper and lower
     strong TDC $\lambda$. Results are obtained with numerical integration on a uniform mesh with $10^{6}$   points.   } 
	\label{tab:Lambda_1} 
\end{table} 
	\newpage
\begin{table}[hbt!] 
\centering
     \scalebox{0.88}{

	\begin{tabular}{ ccccccccc } \hline \\ 
		Copula & Param & $\Lambda^{(l)}_{X|Y}(0.7)$ & $\Lambda^{(l)}_{Y|X}(0.7)$ & $\Lambda^{(u)}_{X|Y}(0.7)$ & $\Lambda^{(u)}_{Y|X}(0.7)$ & $
		\lambda_{l}$ & $\lambda_{u}$ \\ \hline \\ 
		Frechet  &  0.3,  0.7 & 0.06 & 0.06 & 0.06 & 0.06 & 0.30 & 0.3  \\  
		Frechet  &  0.5,  0.5 & 0.19 & 0.19 & 0.19 & 0.19 & 0.50 & 0.5  \\  
		Frechet  &  0.7,  0.3 & 0.43 & 0.43 & 0.43 & 0.43 & 0.70 & 0.7  \\  
		Gumbel   &  1.5  & 0.05 & 0.05 & 0.21 & 0.21 & 0.00 & 0.41  \\  
		Gumbel   &  4  & 0.22 & 0.22 & 0.60 & 0.60 & 0.00 & 0.81  \\  
		Gumbel   &  6  & 0.33 & 0.33 & 0.72 & 0.72 & 0.00 & 0.88  \\  
		Gumbel   &  10 & 0.48 & 0.48 & 0.83 & 0.83 & 0.00 & 0.93  \\   
		Clayton  &  1  & 0.27 & 0.27 & 0.02 & 0.02 & 0.50 & 0.0  \\  
		Clayton  &  2  & 0.44 & 0.44 & 0.04 & 0.04 & 0.71 & 0.0  \\  
		Clayton  &  5  & 0.69 & 0.69 & 0.06 & 0.06 & 0.87 & 0.0  \\  
		Clayton  &  10 & 0.83 & 0.83 & 0.09 & 0.09 & 0.93 & 0.0  \\  
		Clayton  &  30 & 0.94 & 0.94 & 0.17 & 0.17 & 0.98 & 0.0  \\    
		Frank    &  5  & 0.05 & 0.05 & 0.05 & 0.05 & 0.00 & 0.0  \\  
		Frank    &  15 & 0.10 & 0.10 & 0.10 & 0.10 & 0.00 & 0.0  \\    
		Cuadras  &  0.2  & 0.01 & 0.01 & 0.04 & 0.04 & 0.00 & 0.2  \\  
		Cuadras  &  0.5  & 0.03 & 0.03 & 0.21 & 0.21 & 0.00 & 0.5  \\  
		Cuadras  &  0.8  & 0.16 & 0.16 & 0.59 & 0.59 & 0.00 & 0.8  \\     
		Marshall &  0.1,  0.6 & 0.00 & 0.04 & 0.29 & 0.01 & 0.00 & 0.1  \\  
		Marshall &  0.5,  0.5 & 0.03 & 0.03 & 0.21 & 0.21 & 0.00 & 0.5  \\  
		Marshall &  0.6,  0.1 & 0.04 & 0.00 & 0.01 & 0.29 & 0.00 & 0.1  \\  
		Marshall &  0.7,  0.9 & 0.08 & 0.31 & 0.73 & 0.43 & 0.00 & 0.7  \\  
		Gaussian &  0.700 & 0.15 & 0.15 & 0.15 & 0.15 & 0.00 & 0.00  \\  
		Gaussian &  0.900 & 0.32 & 0.32 & 0.32 & 0.32 & 0.00 & 0.00  \\  
		Gaussian &  0.950 & 0.43 & 0.43 & 0.43 & 0.43 & 0.00 & 0.00  \\  
		Gaussian &  0.990 & 0.68 & 0.68 & 0.68 & 0.68 & 0.00 & 0.00  \\ 
		$t$          &  0.500,  2 & 0.16 & 0.16 & 0.16 & 0.16 & 0.42 & 0.42  \\  
		$t$          &  0.900,  2 & 0.46 & 0.46 & 0.46 & 0.46 & 0.73 & 0.73  \\  
		$t$          &  0.950,  2 & 0.58 & 0.58 & 0.58 & 0.58 & 0.81 & 0.81  \\  
		$t$          &  0.500,  1 & 0.20 & 0.20 & 0.20 & 0.20 & 0.56 & 0.56  \\ 
		$t$          &  0.900,  1 & 0.53 & 0.53 & 0.53 & 0.53 & 0.80 & 0.80  \\  
		$t$          &  0.950,  1 & 0.64 & 0.64 & 0.64 & 0.64 & 0.86 & 0.86  \\   
	\end{tabular} 
	}
	        \caption{Values of upper and lower tail dependence measure $\Lambda$ computed for
       the non-GH copulas discussed in Appendix~\ref{AppendixC}. The
       value of the focus parameter is set at $p=0.7$. The last two columns list also the upper and lower
     strong TDC $\lambda$. Results are obtained with numerical integration on a uniform mesh with $10^{6}$   points. } 
	\label{tab:Lambda_2} 
\end{table} 

\newpage
\begin{table}[hbt!]

\centering
		 \scalebox{0.75}{
	\begin{tabular}{ccccccccc}
		&  &  & \multicolumn{3}{c}{$n=500$} & \multicolumn{3}{c}{$n=1000$} \\ \cline{1-9} 
		Distribution & Measure  & Value & mean & std.~dev. & MSE &  mean & std.~dev. & MSE\\ \hline
		\multicolumn{1}{c|}{\multirow{4}{*}{Gumbel}} 
		& \multicolumn{1}{c|}{$\Lambda^{(l)}_{Y|X}$} & \multicolumn{1}{c|}{0.48} &0.484  &0.011  
		& \multicolumn{1}{c|}{$1.241\cdot 10^{-4}$} &0.483  &0.008  &$6.443\cdot 10^{-5}$  \\
		\multicolumn{1}{c|}{} & \multicolumn{1}{c|}{$\Lambda^{(l)}_{X|Y}$} & \multicolumn{1}{c|}{0.48} &0.484  &0.011  
		& \multicolumn{1}{c|}{$1.241\cdot 10^{-4}$} &0.483  &0.008  & $6.443\cdot 10^{-5}$ \\
		\multicolumn{1}{c|}{} & \multicolumn{1}{c|}{$\Lambda^{(u)}_{Y|X}$} & \multicolumn{1}{c|}{0.83} &0.828  &0.0064  
		& \multicolumn{1}{c|}{$4.157\cdot 10^{-4}$} &0.827  &0.0046  &$2.125\cdot 10^{-5}$ \\
		\multicolumn{1}{c|}{} & \multicolumn{1}{c|}{$\Lambda^{(u)}_{X|Y}$} & \multicolumn{1}{c|}{0.83} &0.828  &0.0064  
		& \multicolumn{1}{c|}{$4.157\cdot 10^{-4}$} &0.827  &0.0046  & $2.125\cdot 10^{-5}$ \\ \hline
		\multicolumn{1}{c|}{\multirow{4}{*}{$GH_{2}$}}
		 & \multicolumn{1}{c|}{$\Lambda^{(l)}_{Y|X}$} & \multicolumn{1}{c|}{0.40} &0.372  &0.018  
		 & \multicolumn{1}{c|}{$3.465\cdot 10^{-4}$} &0.392  &0.012  &$1.659 \cdot 10^{-4}$  \\
		\multicolumn{1}{c|}{} & \multicolumn{1}{c|}{$\Lambda^{(l)}_{X|Y}$} & \multicolumn{1}{c|}{0.35} &0.346  &0.017  
		& \multicolumn{1}{c|}{$1.101\cdot 10^{-3}$} &0.351  &0.011  &$2.500 \cdot 10^{-4}$  \\
		\multicolumn{1}{c|}{} & \multicolumn{1}{c|}{$\Lambda^{(u)}_{Y|X}$} 
		& \multicolumn{1}{c|}{0.29} &0.313  &0.019  & \multicolumn{1}{c|}{$2.803\cdot 10^{-4}$} &0.301  &0.011  &$1.433\cdot 10^{-4}$  \\
		\multicolumn{1}{c|}{} & \multicolumn{1}{c|}{$\Lambda^{(u)}_{X|Y}$} & \multicolumn{1}{c|}{0.35} &0.3451  &0.016  
		& \multicolumn{1}{c|}{$2.206\cdot 10^{-3}$}  &0.348  &0.013  &$3.231\cdot 10^{-4}$ \\ \hline
        \multicolumn{1}{c|}{\multirow{4}{*}{t}}
         & \multicolumn{1}{c|}{$\Lambda^{(l)}_{Y|X}$} 
         & \multicolumn{1}{c|}{0.30}   & 0.2994  & 0.0158 & \multicolumn{1}{c|}{$2.501\cdot 10^{-4}$} 
                                      &0.3002  &0.0119  &$ 1.413 \cdot  10^{-4}$  \\
         \multicolumn{1}{c|}{} & \multicolumn{1}{c|}{$\Lambda^{(l)}_{X|Y}$} 
          & \multicolumn{1}{c|}{0.30}  & 0.2994  & 0.0158  & \multicolumn{1}{c|}{$2.501\cdot 10^{-4}$}
                                      &0.3002  &0.0119 &$ 1.413\cdot  10^{-4}$  \\
         \multicolumn{1}{c|}{} & \multicolumn{1}{c|}{$\Lambda^{(u)}_{Y|X}$}
          & \multicolumn{1}{c|}{0.30}  & 0.2994  & 0.0157  & \multicolumn{1}{c|}{$2.495\cdot 10^{-4}$} 
                                      &0.3004  &0.0119  &$1.433\cdot 10^{-4}$  \\
         \multicolumn{1}{c|}{} & \multicolumn{1}{c|}{$\Lambda^{(u)}_{X|Y}$} 
          & \multicolumn{1}{c|}{0.30}  & 0.2994  & 0.0157   & \multicolumn{1}{c|}{$2.495\cdot 10^{-4}$} 
                                      &0.3004  &0.0119  &$1.433\cdot 10^{-4}$ \\ \hline
	\end{tabular}
	}
	
	\caption{Performance of MLE estimators of upper and lower
          measures of tail dependence, $\Lambda^{(u)}$ and
          $\Lambda^{(l)}$, obtained by simulating from three different
          Copulas: a Gumbel with parameter $\theta=10$, a GH
          distribution with parameters specified in the secon row of
          Table \ref{tab:GH_Table}, and a $t$ distribution with
          parameters $\rho=0.8$ and $\nu=3$. Summary measures reported
          over
          1,000 replicatios for each of two sample sizes are: the
          mean, standard deviation (std.~dev.), and mean squared error (MSE).}
	\label{tab:sim} 

\end{table}

\bibliographystyle{acm}
\bibliography{TailDep_arxiv}

\end{document}